\documentclass[aps,prc,twocolumn,floatfix,showpacs,a4paper,
nofootinbib,amsmath,amssymb]{revtex4-1}
\usepackage{graphicx}
\usepackage{dcolumn}
\usepackage{bm}
\usepackage{color}

\newcommand{\be}{\begin{equation}}
\newcommand{\ee}{\end{equation}}
\newcommand{\ba}{\begin{eqnarray}}
\newcommand{\ea}{\end{eqnarray}}
\newcommand{\bd}{\begin{displaymath}}
\newcommand{\ed}{\end{displaymath}}
\newcommand{\bea}{\begin{eqnarray}}
\newcommand{\eea}{\end{eqnarray}}

\renewcommand{\vec}[1]{\mbox{\boldmath$#1$}}
\DeclareMathOperator{\Artanh}{Artanh}
\DeclareMathOperator{\Arsinh}{Arsinh}

\begin{document}
\title{An Initial State with Shear in Peripheral Heavy Ion Collisions}

\author{V.K. Magas$^1$, J. Gordillo$^1$, D. Strottman$^{2,3}$, 
Y.L. Xie$^{2,4}$ and L.P. Csernai$^2$}
\smallskip

\affiliation{
$^1$Departament 
Universitat\! de\! Barcelona, 08028 Barcelona, Spain\\
$^2$Institute of Physics and Technology, University of Bergen,
Allegaten 55, 5007 Bergen, Norway\\
$^3$Los Alamos National Laboratory, Los Alamos, 87545 New Mexico, USA\\
$^4$Frankfurt Institute for Advanced Studies,
Ruth-Moufang-Str. 1, 60438 Frankfurt am Main, Germany
}

\begin{abstract}
In the present work we propose a new initial state model for hydrodynamic simulation of relativistic heavy ion collisions based on Bjorken-like solution applied streak by streak in the transverse plane. 
Previous fluid dynamical calculations in Cartesian coordinates with an initial state based on a streak by streak Yang-Mills field led for peripheral higher energy collisions  to large angular momentum, initial shear flow and significant local vorticity. Recent experiments verified the existence of this vorticity via the resulting polarization of emitted $\Lambda$ and $\bar{\Lambda}$ particles. At the same time parton cascade models indicated the existence of more compact initial state configurations, which we are going to simulate in our approach. 

The proposed model satisfies all the conservation laws including  conservation of a strong initial angular momentum which is present in non-central collisions. As a consequence of this large initial angular momentum we observe the rotation of the whole system as well as the fluid shear in the initial state, which leads to large flow vorticity. Another advantage of the proposed model is that the initial state can be given in both [t,x,y,z] and $[\tau, x, y, \eta]$ coordinates, and thus can be tested by all 3+1D hydrodynamical codes which exist in the field. 

\end{abstract}
\date{\today}


\maketitle

\section{Introduction}

About 15 years ago a nucleus-nucleus Initial State (IS) model was constructed  \cite{M2001,M2002}  based on the longitudinal effective string rope model for realistic, 3+1D relativistic fluid dynamical models. This model  preceded the early development of Quark-Gluon Plasma (QGP) research, but still it reflected correctly not only the energy-momentum, but also angular momentum conservation, initial shear flow, and local vorticity. Actually as a consequence of the large initial angular momentum present in the non-central ultra-relativistic heavy ion collisions, the effective rotation of the whole fireball has been observed once the effective string rope model was applied to simulate Pb+Pb collisions at ALICE@LHC \cite{CM_v1}.  Obviously such a rotation leads to large flow vorticity \cite{CMW13}.

Several other initial state models neglected these basic features, but recent experimental and theoretical developments indicate that angular momentum, local vorticity and the subsequent particle polarization is observable and provides valuable information about the QGP. Recently, significant $\Lambda$ polarization was detected and analyzed in detail in the RHIC BES program \cite{Lisa,Nature}.  These results indicate that shear and vorticity should not be neglected if we wish to account for the observed polarization.

\begin{figure}[ht]     
\begin{center}
\resizebox{0.98\columnwidth}{!}
{\includegraphics{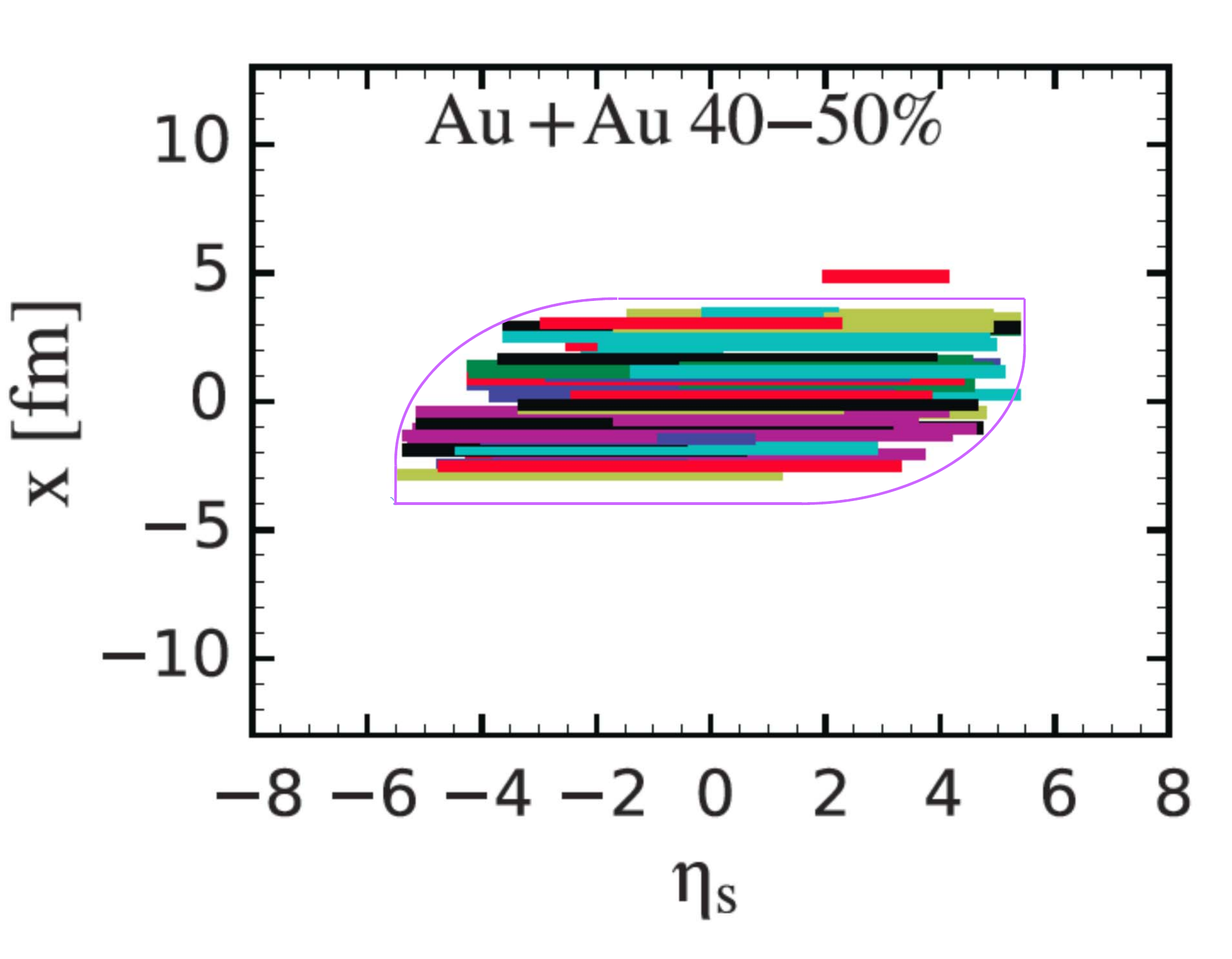}}
\caption{ (color online)
Initial State streaks indicate the energy density distribution with
fluctuating initial conditions from A Multi-Phase Transport (AMPT) 
model \cite{Fig.1}. The model simulates a
 Au+Au collisions at  the energy $\sqrt{s_{NN}}=200 $ GeV of centrality percentage 40-50\% in the reaction plane in $[x,\eta_s]$ coordinates. The energy density distribution remains compact and the off-center side streaks are actually shorter than the central streaks.At the same time the off-center side streaks have obviously moved forward or backward indicating angular momentum conservation and significant local shear. The inserted thin magenta line surrounding the matter distribution shows the characteristic shape of the initial state obtained in the AMPT model.}
\label{Streaks}
\end{center}
\end{figure}

Several parton kinetic and field theoretical models were recently implemented to describe the IS, providing a rather different initial state configuration, especially for non-central collisions \cite{Fig.1,PangEA2016,Pathia,AMPT,Deng2016,Vovchenko}.  While in  peripheral collisions \cite{M2001,M2002} the off-center streaks were assumed to have relatively weak fields and therefore showed large longitudinal extent, the present kinetic models show a more compact IS, where the streaks away from the center are more compact and experience stronger fields. We can see this in Fig. \ref{Streaks}.

This figure is a result of the simulation of Au+Au collisions at  the energy $\sqrt{s_{NN}}=200 $ GeV of 40-50\% centrality  by means of A Multi-Phase Transport (AMPT) model \cite{Fig.1}.  As we can see this is an example of such an initial state configuration,  which is more compact than the afore mentioned early initial  state models \cite{M2001,M2002}. 
Note that the off-center side streaks are actually shorter than the central streaks, and at the same time they have obviously moved forward and backward indicating angular momentum conservation and significant local shear. The thin magenta  external contour line, inserted in Fig. \ref{Streaks}, surrounding the matter distribution, shows the characteristic shape of the initial state obtained in the AMPT model, which provides  us a guidance to form  a more compact initial state model   with fixed longitudinal extent of the projectile/target side   peripheral streaks. 
   
Those models that account for the initial shear and vorticity \cite{CMW13,Bec13,Erratum,WCBS14,PangEA2016} could predict  and study the observed polarization.

These developments make it timely that in view of new experimental 
and theoretical developments we need to revisit the early IS model,
with the aim of keeping all basic features as local shear, angular
momentum conservation and local vorticity, while adapting to the
developments in parton kinetic \cite{Pathia,AMPT,Deng2016} and 
field dominance \cite{Vovchenko} models. Furthermore, as 
several field theoretical models have been developed recently in the
proper-time and space-time rapidity, Milne coordinates 
$[\tau, x, y, \eta]$, we also present the model in the
same way to make it useful for other approaches. On the other
hand we will continue to use fluid dynamical models in 
Cartesian coordinates, $[t,x,y,z]$, as {\it e.g.}, the Particles in Cell Relativistic 
model (PICR), since in these codes
the numerical effects are well studied.

\section{Heavy ion collisions as a set of independent streak-streak collisions}

Let us consider a peripheral heavy ion collision at highly relativistic energies. The projectile and target are strongly Lorentz contracted before the collision, while the parton momentum distributions of the projectile and target are strongly Lorentz elongated.

We divide up the transverse plane into cells of less than 1 fm$^2$ size. The corresponding elements of the projectile and target hit and inter-penetrate each other. The transverse expansion can be ignored for the first moments of the collision, and thus, at the beginning, one can describe the nucleus-nucleus collision as a set of independent streak-streak collisions, corresponding to the same transverse coordinates. One or two fm/c after the first contact the partons from the pre collision projectile and target slabs will form a streak, which is about two to four fm long. 

Throughout this paper we shall assume that the projectile will have an initial positive (forward) momentum and the target will have a negative momentum in the c.m. system.  

Due to the large momentum spread of the initial partons the resulting streaks will have a mixture of projectile and target partons at each point of the streak.  In peripheral collisions in a given slab-slab collision there will be a projectile/target asymmetry, except at the central streaks corresponding to $x=x_c$. The final streaks will have a finite longitudinal momentum, which can be calculated using the longitudinal momentum conservation from momenta of the two original slabs. Streaks on the projectile side of the transverse plane ($x>x_c$) will have a forward momentum, while on the target side ($x<x_c$) a backward one. Thus, each streak will have its own center-of-mass (c.m.) reference  system.  Neighboring streaks may have an initial shear. 

This compact 
system will have initially ($\sim 1{-}2$ fm/c) a non-zero angular momentum.
Its partons will be mixed from the projectile and target. The 
chromo-magnetic forces (string tension) will attract the leading
partons.  So, the system will not expand with the speed of light 
but will be held back by the fields. The original Lorentz
elongation of the momentum distribution and the field attraction 
will lead to an initial parton distribution, which will be close 
to uniform, as both the target and projectile partons can 
populate the whole length of the moving streak \cite{Vovchenko}, as assumed in the 
Bjorken model. 

Let us consider projectile and target slabs colliding head on with each other at a given transverse point $[x_i,y_i]$.  The main ansatz of this work is to assume that the Bjorken model can be applied to describe these slab-slab collisions during the initial stage of the reaction. This means that  the resulting streak of matter has a longitudinal rapidity profile as in the Bjorken flow expansion, contrary to the rapidity profiles used in \cite{M2001,M2002,MiL2012,Mishustin2011}. However, each of these streaks will be described by the Bjorken flow in its own c.m. frame. The overall reaction volume, {\it i.e.}, all the streaks together, can be described in the Lab frame, where for each such streak $i$ we can construct, as we shall see later,  a starting point $[t_{i0},z_{i0}]$.

From the initial geometry we know for each final streak $i$, at a transverse point $[x_i,y_i]$, 
the total baryon charge, the total kinetic energy, and the total momentum in the longitudinal direction.   {\it For simplicity we will drop the subindex $i$ in the rest of this section, since here we will only be interested in one transverse position. Later, in order to describe the whole collision, all the quantities introduced and derived in this section will have the subindex of the corresponding transverse position. }
 
\begin{figure}[ht]     
\begin{center}
\resizebox{0.9\columnwidth}{!}
{\includegraphics{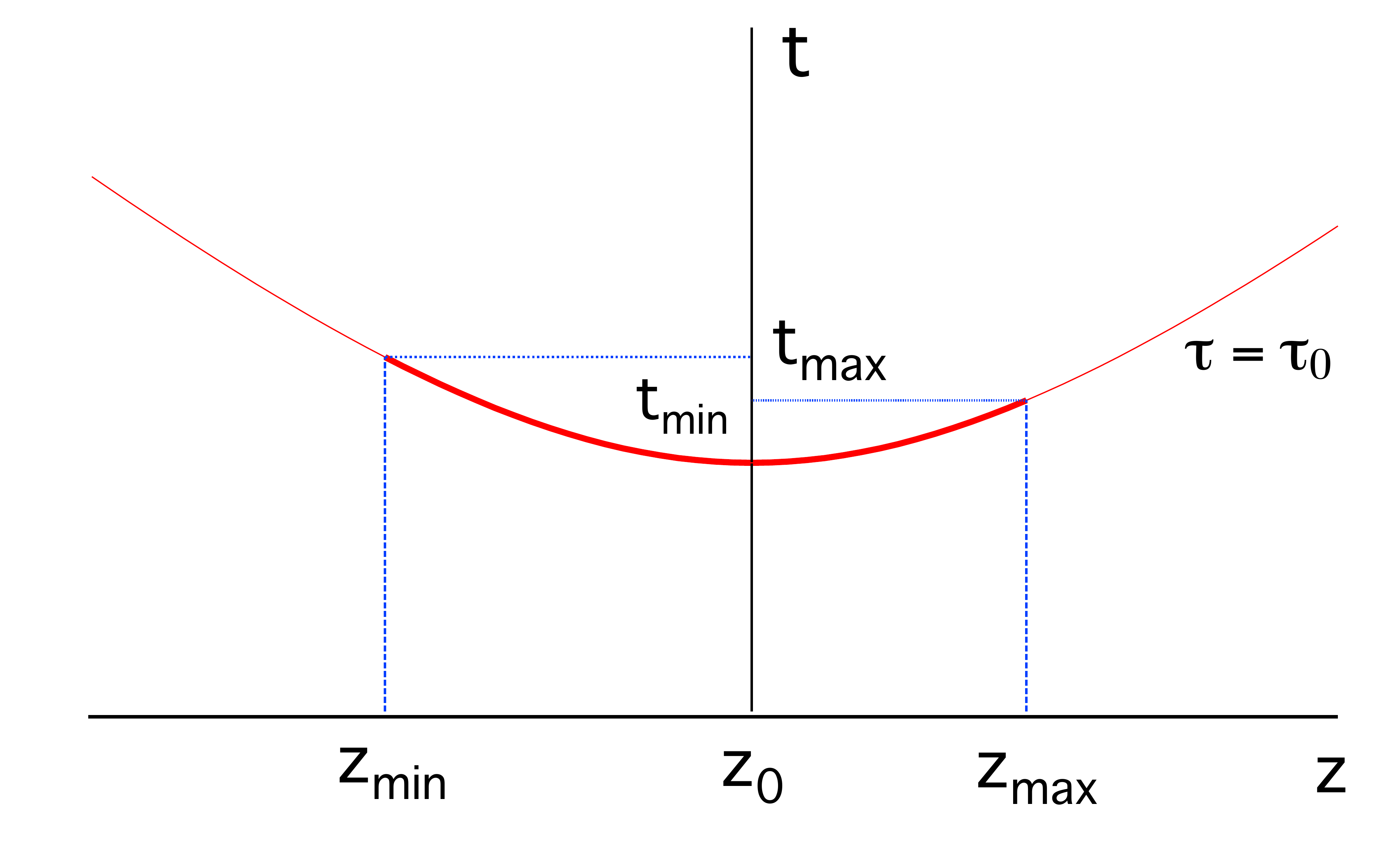}}
\caption{ (color online)
Space-time sketch of a slab-slab collision in the collider lab frame.  At proper time $\tau_0$ after the collision, the resulting streak stretches from $t_{min}, z_{min}$  to $t_{max}, z_{max}$.
}
\label{f1}
\end{center}
\end{figure}

Thus, we want to describe collisions of two slabs of the nuclear matter with 
$N_1, E_1, P_{1z}$ and $N_2, E_2, P_{2z}$ for projectile and target respectively. The pre-collision projectile slab moves with the beam rapidity $y_0$ while the target slab
with $-y_0$. In Fig. \ref{f1} we show the streak-streak collision  (asymmetric in the general case). The first contact happens at $(t_0,z_0)$, and at the proper time $\tau_0$ the resulting streak stretches from $t_{min}, z_{min}$ 
to $t_{max}, z_{max}$.

The transformation between the usual Cartesian coordinates $x^\mu =(t,x,y,z)$ and Milne coordinates
$\tilde{x}^\mu =(\tau,x,y,\eta)$  is given as
\ba
t - t_0 &=& \tau \cosh \eta  \ ,
\nonumber \\
z - z_0 &=& \tau \sinh \eta \ ,
\label{tz1}
\ea
\ba
\tau &=& \sqrt{(t-t_0)^2 - (z-z_0)^2} \ ,
\nonumber \\
\eta &=& \frac{1}{2} \ln\left(\frac{t-t_0 + z-z_0}{t-t_0 - (z-z_0)}\right) = \Artanh \frac{z-z_0}{t-t_0} \  ,
\label{taueta1}
\ea
where $\eta$ is the space-time rapidity. 
Consequently,
\ba
dt &=& \cosh \eta\, d\tau + \tau \sinh \eta\, d\eta \ ,
\nonumber \\
dz &=& \sinh \eta\, d\tau + \tau \cosh \eta\, d\eta \ ,
\label{dtdz1}
\ea
\ba
dz\, dt &=& \tau d\eta d\tau\ .
\label{dtdz2}
\ea
In case of the longitudinal Bjorken scaling flow, 
the local flow velocity of matter is 
\be
u^\mu ={x^\mu}/{\tau}= (\cosh \eta, 0,0\, \sinh \eta)\,.
\label{bjflow}
\ee
Thus the velocity of the Bjorken flow at point ($t,z$) is
\be
v_z = \frac{z-z_0}{t-t_0} \,,   
\ee
and for the streak ends we can write
\ba
\eta_{max}  & = & \Artanh v_{max} = \Artanh \frac{z_{max}-z_0}{t_{max}-t_0}\,, \nonumber \\
\eta_{min}  & = & \Artanh v_{min} = \Artanh \frac{z_{min}-z_0}{t_{min}-t_0} \,,
\label{etaMM}
\ea
or in other words 
\ba
z_{max} & = & z_0 + \tau_0 \sinh \eta_{max}\,, \nonumber \\
z_{min} & = & z_0 + \tau_0 \sinh \eta_{min}  \,.
\label{zMM}
\ea

\subsection{Conservation Laws}

Following the philosophy of the Bjorken model we assume that each streak at the moment when its proper time $\tau$ is equal to $\tau_0$,  contributes to the initial state at local thermal equilibrium. Then it evolves 
further according to the hydrodynamic Bjorken equations. The main 
characteristic of this $\tau=\tau_0$ state is that it is constant as a function of $\eta$, while the local flow four-velocity is given by eq. (\ref{bjflow}).
The initial energy and baryon densities can be found from the conservation laws.

The $\tau=const.$ hypersurface normal four vector is given as,
\be
d^3\Sigma_{\mu}^{( t,z)}=\tau  \left( \cosh \eta 
,0 ,0,-\sinh \eta\right) dxdy d\eta = \tau A u_\mu d\eta,
\ee
where $A$ is the transverse cross section of the 
streak (in the $[x,y]$-plane).

The net baryon four current for a streak is
$N^\mu = n u^\mu$, and thus the net baryon number crossing 
a constant $\tau$- hypersurface element is
\be
dN = d^3\Sigma_\mu N^\mu =  n \tau A \, d\eta , 
\ee

\begin{figure}[ht]     
\begin{center}
\resizebox{1.01\columnwidth}{!}
{\includegraphics{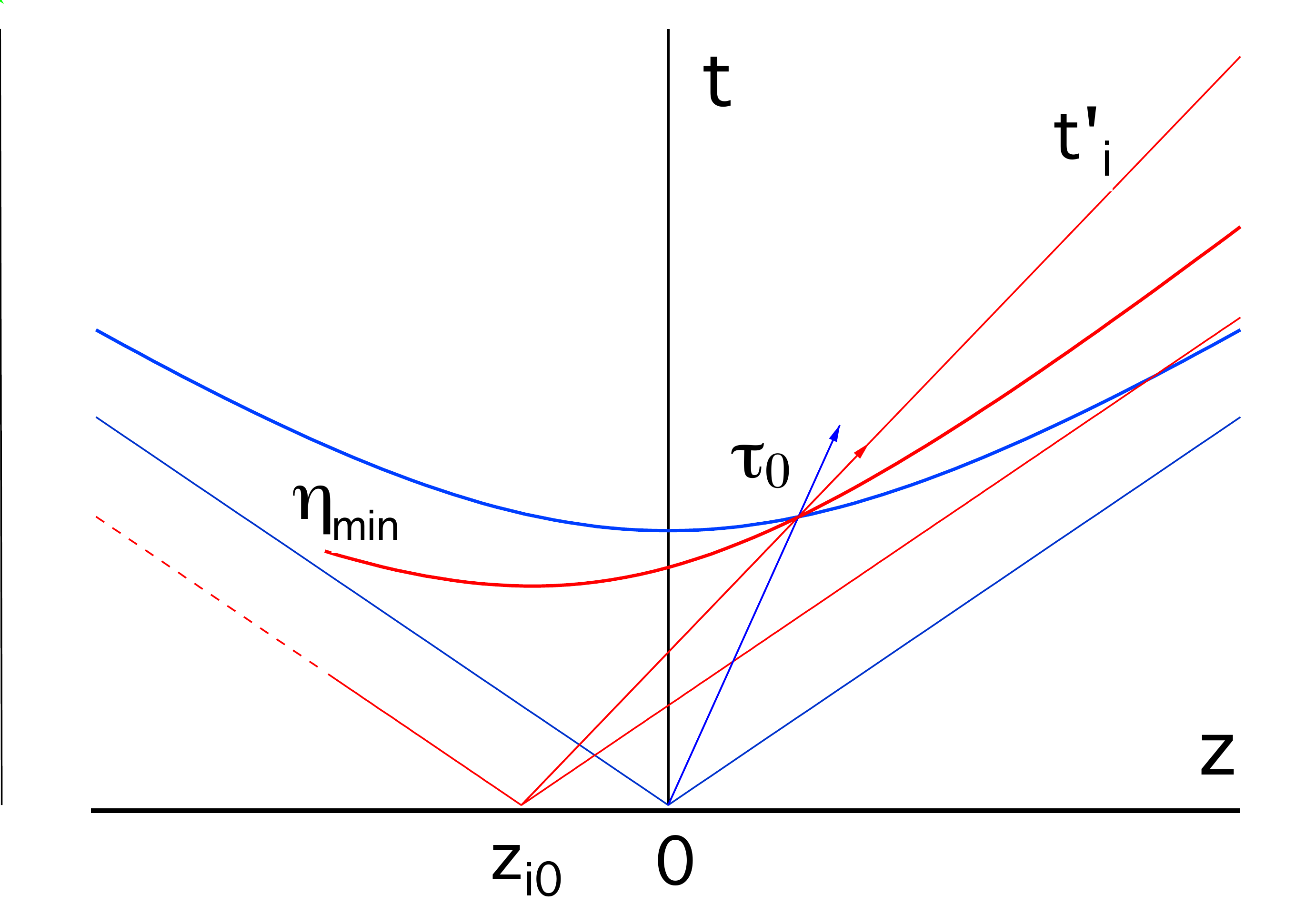}}
\caption{ (color online)
Space-time sketch of the central streak (blue), 
and the $i$-th peripheral streak (red)
on the projectile side. 
The origins of the streaks are not identical, but at the
proper time, $\tau_0$ and their leading edge position, $z_{max}$ match.
The two streaks cross each other in the space-time at the $t'_i$ axis where
both streaks have the same proper time $\tau_0$. 
This axis corresponds to the c.m. rapidity of the $i$-th peripheral
streak, $\eta_i$.
We can see that the
local four velocity vectors are different for the two streaks causing shear
and vorticity.
}
\label{f2}
\end{center}
\end{figure}        
%

Thus, the baryon number conservation for a streak, 
assuming uniform $\eta$-distribution gives:
\be
N = N_1 + N_2 = \tau_0 n(\tau_0) A \left[\eta_{max} - \eta_{min}\right]\,  ,
\label{Ncons}
\ee
where ($N_1 + N_2$) is an invariant scalar given by the 
Projectile (1) and Target (2) baryon charge contribution to a given streak,
and the 
difference, $(\Delta\eta = \eta_{max} - \eta_{min})$, should also be a 
boost invariant quantity.

The energy-momentum tensor is 
$T^{\mu \nu} = e u^\mu u^\nu - p\Delta^{\mu \nu} + \pi^{\mu \nu}$, where $\Delta^{\mu \nu} = g^{\mu \nu} - u^\mu u^\nu$ is the 
projection tensor and $\pi^{\mu \nu}$ is the
shear-stress tensor, both orthogonal to the flow velocity.
Energy crossing the $\tau=const.$ hypersurface element is 
\ba \nonumber
dE &=& d^3\Sigma_\mu T^{0\mu} 
= \tau A \, [e u^0 u^\mu  - p \Delta^{0\mu} + \pi^{\mu 0}] u_\mu d\eta \,  
\\
&=& \tau A\, e \, u^0 \, d\eta = \tau A\, e\, \cosh \eta\, d\eta .
\ea
Integrating this between $\eta_{max}$ and $\eta_{min}$, leads to
\be
E = E_1 + E_2 = \tau_0 e(\tau_0) A (\sinh \eta_{max} - \sinh \eta_{min}) \, .
\label{E12}
\ee
Note that both $E_1 + E_2$ and ($\sinh \eta_{max} - \sinh \eta_{min}$) are 
frame dependent. Nevertheless, the equations for $N$ and $E$ have
the same form in any boosted frame.

\begin{figure}[htb]     
\begin{center}
\resizebox{1.01\columnwidth}{!}
{\includegraphics{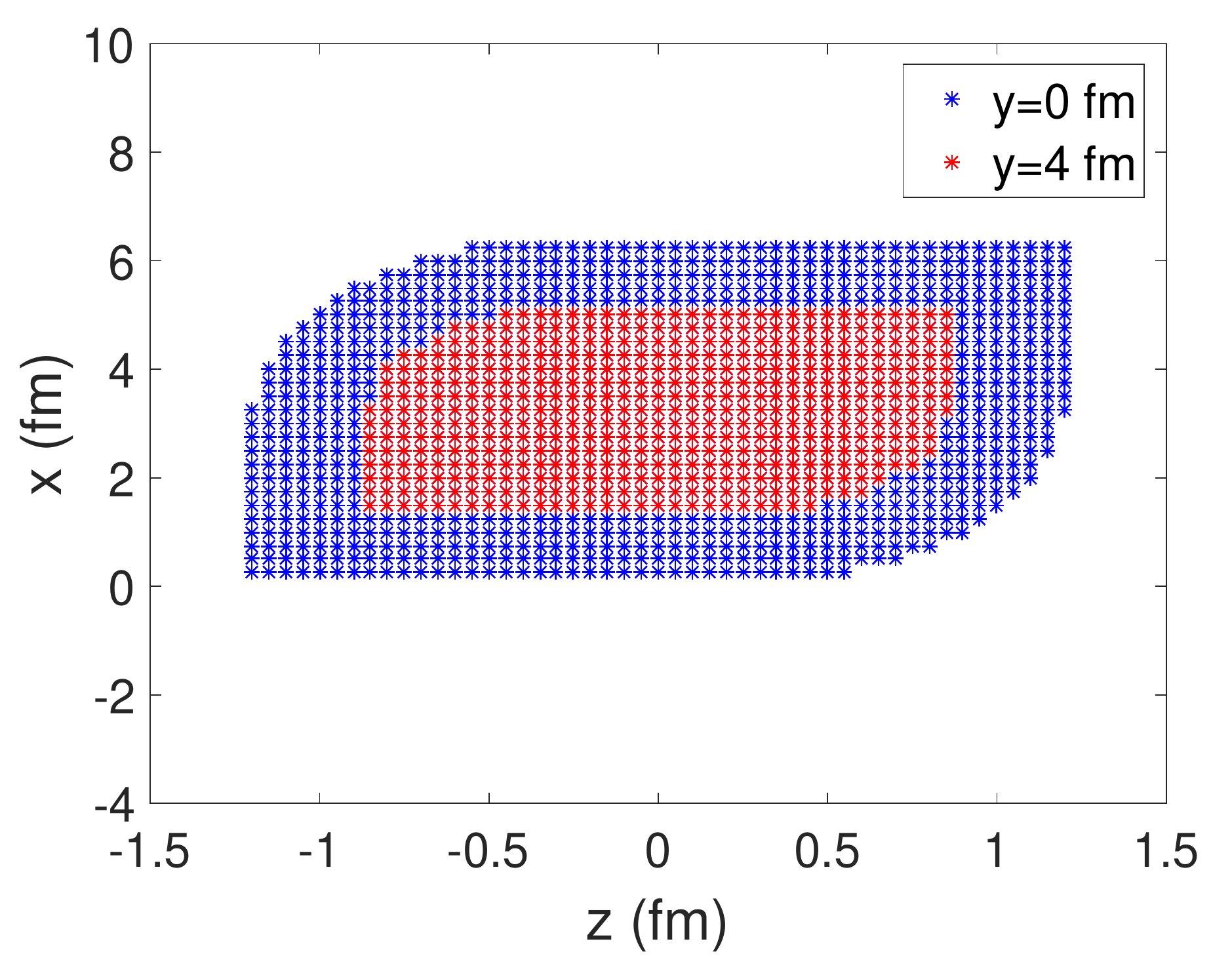}}
\caption{ (color online)
The initial configuration of the streaks in the reaction plane, 
on the $[x,\eta, \tau_i=\tau_0=1{\rm fm/c}]$-hypersurface for $y=0$
(blue streaks) and for $y=4$ fm (red streaks) overlaid.
An assumption of our model is that the streak energy density is uniform and is the same for all streaks.
The configuration is qualitatively similar to the parton cascade result
shown in Fig. \ref{Streaks}. This example is calculated for a
Au+Au reaction at $100+100$ GeV/nucl energy and impact parameter
$b= 0.5 (R_{Pb} + R_{Pb}) = 6.5$ fm, correspondingly the $y=0$ plane crosses the $x$-axis at $x_c=3.25$ fm. 
We have fixed our model parameters as $\tau_0 = 1.0$ fm/c and $\Delta \eta_c = 2$, 
which leads to the energy density $e_i(\tau_0)=e_c(\tau_0)  = 156.31$ GeV/fm$^3$. 
Subsequent figures were calculated with the same reaction
parameters.  Note that this figure serves only for a qualitative 
understanding of the model, since each streak is plotted at the moment when its $\tau_i=\tau_0$.
}
\label{Config-in-xz}
\end{center}
\end{figure}  

Similarly for the longitudinal momentum component we have 
\be
d P_z =  d^3\Sigma_\mu T^{z\mu}
\ee
and it follows that
\ba \nonumber
P_z &=& \tau_0 A\, \int [e u^z u^\mu  - p \Delta^{z\mu} + \pi^{\mu z}]
u_\mu d\eta \\
&=& \tau_0 A\, \int e\, u^z\, d\eta = 
\tau_0 A\, e\, \int \sinh \eta\, d\eta,
\label{Pzeta}
\ea
and so 
\be
P_{z} = 
P_{1z} - P_{2z} = \tau_0 A\, e\, (\cosh \eta_{max} - \cosh \eta_{min})\ .
\label{P13}
\ee

\begin{figure}[htb]     
\begin{center}
\resizebox{0.94\columnwidth}{!}
{\includegraphics{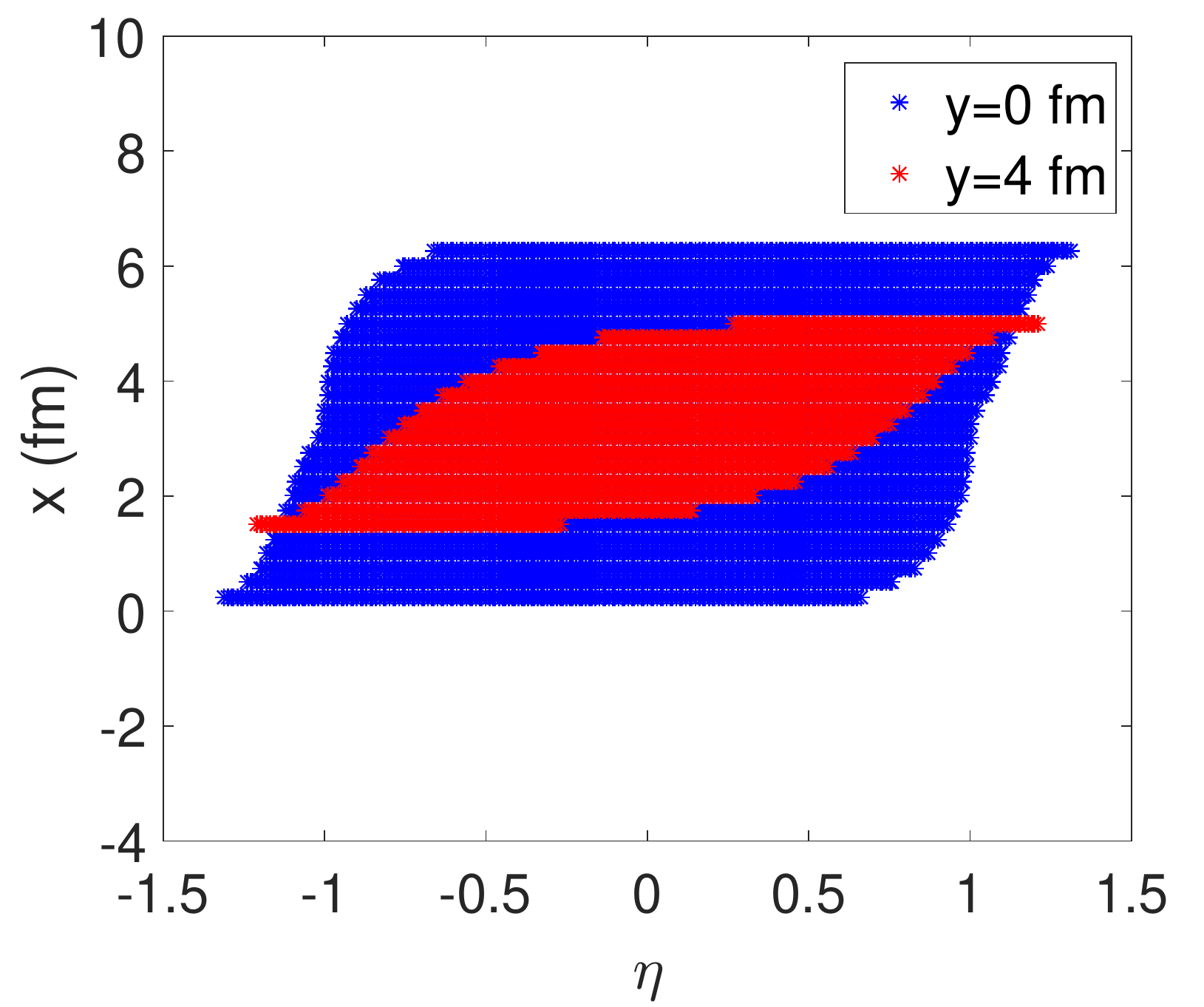}}
\caption{ (color online)
The initial configuration of the streaks in the reaction plane, 
on the $[x,\eta, \tau_i=\tau_0=1{\rm fm/c}]$-hypersurface,
plotted versus the geometrical rapidity $\eta$.
Shown are the layers at $y=0$ (blue streaks) and at 
$y=4$ fm (red streaks) overlaid.
The streak energy density is uniform in $\eta$.
The obtained configuration is also qualitatively similar to the parton cascade 
result shown in Fig. \ref{Streaks}, but the top and bottom
edges show a special behaviour. This example is calculated for the
same parameters as Fig. \ref{Config-in-xz}. Note that this figure serves only for a qualitative  understanding of the model, since each streak is plotted at the moment when its $\tau_i=\tau_0$.
}
\label{Config-in-xeta}
\end{center}
\end{figure}  

The above equations can be given in a more compact form if we introduce for each streak instead of $\eta_{max}$ and  $\eta_{min}$ two other quantities, namely
semi-difference $\Delta \eta/2$ and c.m. rapidity $<\eta>$, given as
\be
\frac{1}{2}\Delta \eta = 
\frac{\eta_{max}-\eta_{min}}{2}  ,
\ee
\be
<\eta > = 
\frac{\eta_{max}+\eta_{min}}{2} .
\ee 
With these parameters, from eqs. (\ref{Ncons},\ref{E12},\ref{P13}) 
it follows that 
\be
N = \tau_0 n(\tau_0) A \Delta \eta \,,
\label{N_eta}
\ee
\be
E = 2 \tau_0 e(\tau_0) A \sinh(\Delta \eta/2) \cosh(<\eta>)
\,,
\label{E_eta}
\ee
\be
P_{z} = 2 \tau_0 e(\tau_0) A \sinh(\Delta \eta/2) \sinh(<\eta>)
\,. 
\label{P_eta}
\ee

Comparing eqs.  (\ref{E_eta}) and  (\ref{P_eta}) we find an expression for c.m. rapidity:
\be
<\eta> = 
\Artanh \frac{P_{z}}{E}\ .
\label{eta}
\ee

\begin{figure}[htb]     
\begin{center}
\resizebox{1.01\columnwidth}{!}
{\includegraphics{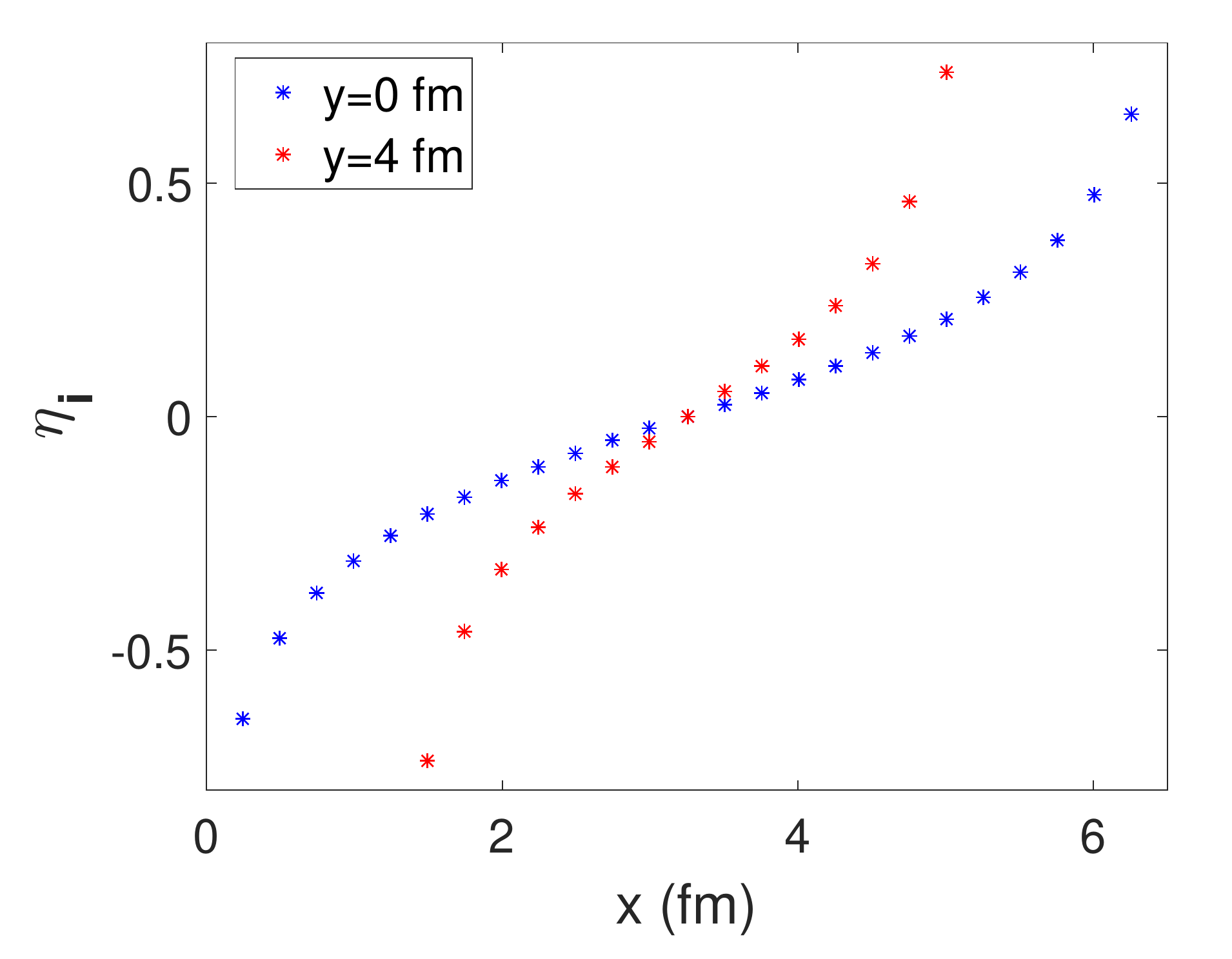}}
\caption{ (color online)
The mid rapidity of the streaks $<\eta_i>$ in the reaction 
plane $[x,z]$  for different values of $x$, for $y=0$
(blue stars) and for $y=4$ fm (red stars),
for the same reaction and parameters listed in Fig. \ref{Config-in-xz}.
}
\label{eta-i}
\end{center}
\end{figure}        
%

\section{The implementation of the model}
\label{sec3}

One must select an initial proper time parameter, $\tau_0$,  for the model
which  can be chosen relatively freely.
In the literature the typical values of the Bjorken initial proper time $\tau_0$
vary from $0.1$ fm to a few fm. 
According to our assumptions the fluid elements  show a Bjorken-type  scaling expansion where the flow rapidity equals the rapidity coordinate, $\eta$,
of a given fluid element of the streak in the rest frame
of the streak (RFS).
The streaks corresponding to different transverse coordinates, $[x_i,y_i]$, have in general different reference rest frames, RFS$_i$, and different
initial points $t_{i0}$ and $z_{i0}$.

\begin{figure}[htb]     
\begin{center}
\resizebox{1.01\columnwidth}{!}
{\includegraphics{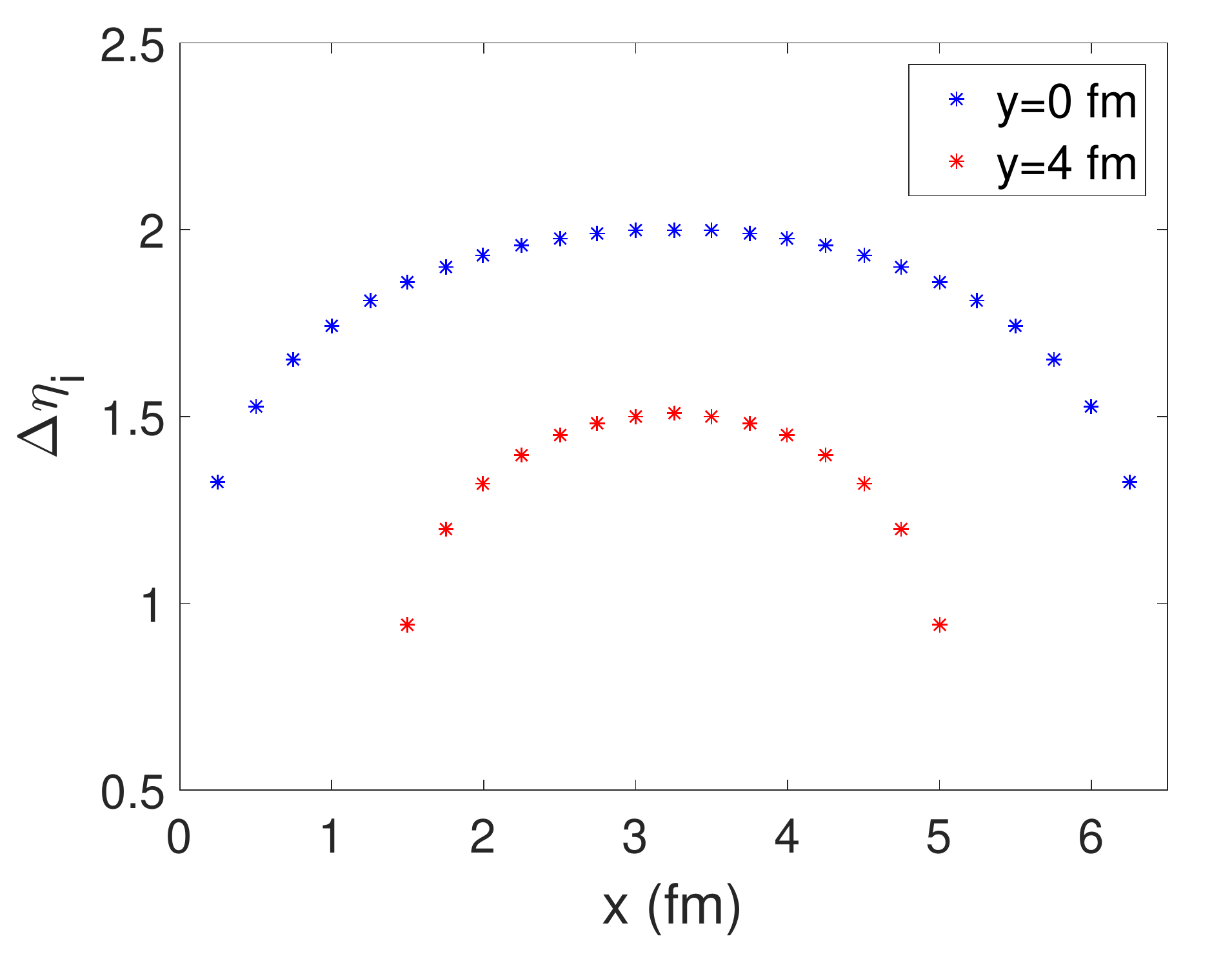}}  
\caption{ (color online)
The rapidity spread of the streaks $\Delta\eta_i$ in the reaction 
plane $[x,z]$ for $y=0$
(blue stars) and for $y=4$ fm (red stars),
for the same reaction and parameters listed in Fig. \ref{Config-in-xz}.
}
\label{Delta-eta}
\end{center}
\end{figure}        

{\bf Central streak:}  

Let us  now consider the central streak, denoted by index $i = c$.
For this streak the collider c.m. frame and the rest frame
of the streak (RFS) are the same, and correspondingly $\tau_0 = t_0$.
The total energy of this streak is then $E_c$, 
and its momentum  is $P_{cz}=0$, and correspondingly, see eq. (\ref{eta}),  $<\eta_c>=0$.
Thus for the central streak we have $-\Delta \eta_c /2 <\eta_c < \Delta \eta_c /2$.

We choose also a coordinate system so that the first contact of the target and projectile slabs for the central streak happens at $t=0$ and $z=0$. Thus, for the central streak 
the starting coordinates for the Bjorken solution are $t_{0,c} = 0$ and $z_{0,c} = 0$, while these have to be calculated for the peripheral streaks.

Then at $\tau_c=\tau_0$ the length of the central streak 
is $\Delta z_c = z_{c-max} - z_{c-min}= 2 \tau_0 \sinh \left(\Delta \eta_{c} /2\right)$.  
The extension of the central streak  in the geometrical rapidity space, $\Delta \eta_{c}$, is one of our free parameters.

The energy density of the central streak at $\tau_c=\tau_0$ is given by eq. (\ref{E12}):
\be
e_c(\tau_0) =
E_c \left/\left[ 2\, \tau_0\, A\, \sinh \left(
\Delta\eta_c/2\right) \right] \right. \ .
\label{EC}
\ee

Also, once $\tau_0$ and $\Delta \eta_c/2$ are set, using
eq. (\ref{tz1}) we get the position of the forward edge of the
central streak at 
\ba
z_{max}^{c} &=& \tau_0 sinh(\Delta \eta_c /2) \ \ ,
\nonumber \\
t_{max}^{c} &=& \tau_0 \cosh (\Delta\eta_c /2 ) \ \ .
\label{ztcm}
\ea
The position of the back edge can be calculated the same way,
$t_{min}^{c}=t_{max}^{c}$ and $z_{min}^{c}=-z_{max}^{c}$.
\smallskip

{\bf Peripheral streaks:}  

At finite impact parameter the asymmetry of the projectile and target
side leads to a finite momentum, $P_{iz}$, for the peripheral streak $i$. 
Unlike in the usual approach, we do not set the origin of all streak hyperbolae
to the same point (as the central streak); instead we make two  assumptions:

{\bf(a)} that in the collider c.m. frame the
leading edge of the Projectile (P) side streaks will be aligned uniformly at the moment $\tau_i=\tau_0$, 
{\it i.e.}, the $z_{max}^{i,P}$
values will be the same on the  projectile side, for all $i,P$-s 
\be
z_{max}^{i,P}=z_{max}^{c}
\ee
(and on the Target (T) side, for all $i,T$-s the $z_{min}^{i,T}=z_{min}^c$).
This reflects the 
behavior of the parton kinetic models as shown in Fig. \ref{Streaks}.

\begin{figure}[htb]     
\begin{center}
\resizebox{1.01\columnwidth}{!}
{\includegraphics{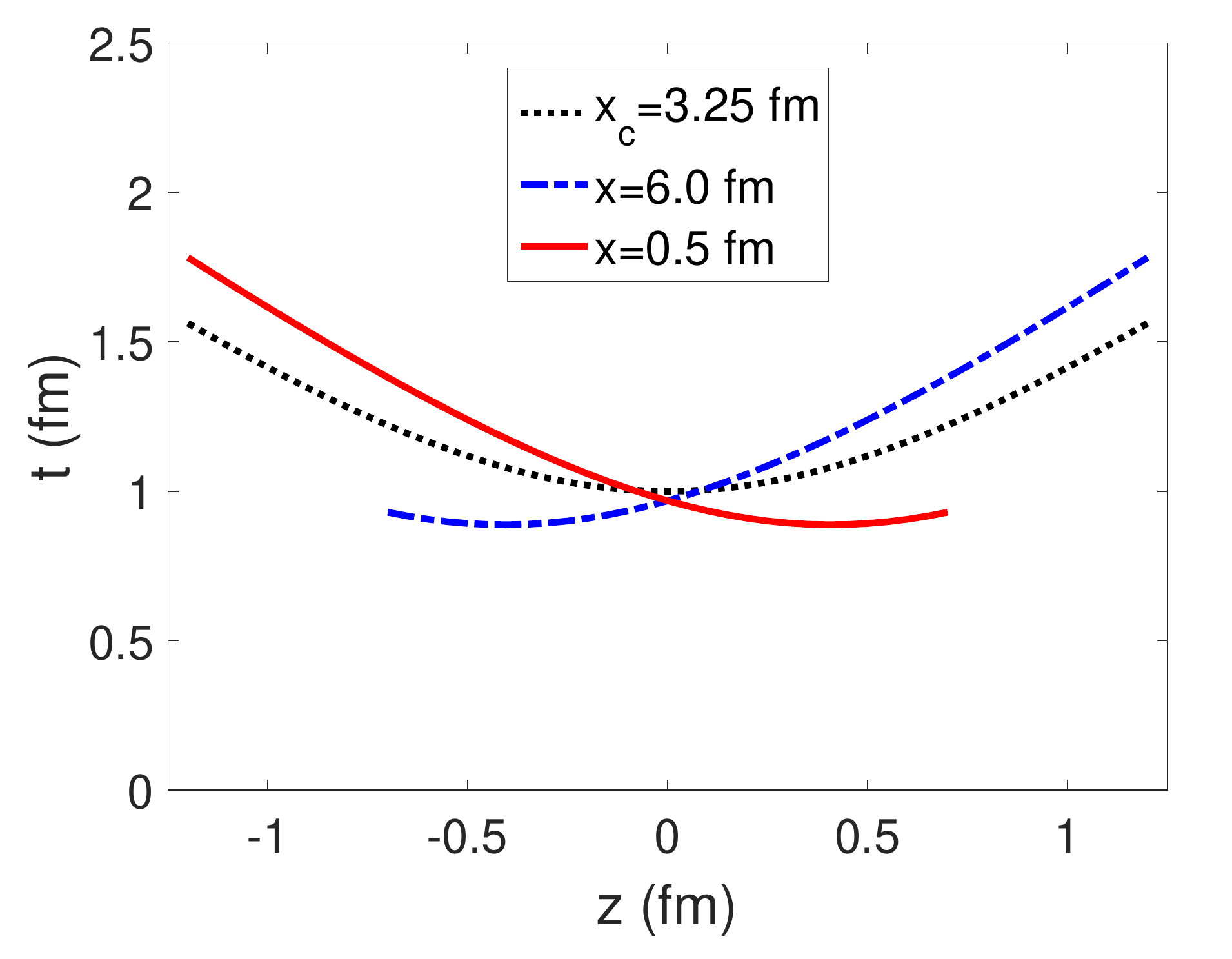}}
\caption{ (color online)
Space-time, $[t,z]$ configuration of the central streak, 
and two peripheral streaks,
on the projectile and target sides. 
The origins of the streaks are not identical but at the
proper time, $\tau_0$  the leading (P) and trailing (T) edge positions, 
$z_{max}, z_{min}$ match for the two side streaks, respectively.
The two streaks cross each other in the space-time at the points where
both streaks have the same proper time $\tau_0$.
The reaction parameters are the same as 
listed in Fig. \ref{Config-in-xz}.
}
\label{st-2}
\end{center}
\end{figure}        
%

%

%
{\bf (b)}
that at the moment $\tau_i=\tau_0$ in the corresponding RFS$_i$
\be
e_i(\tau_0) = e_c(\tau_0) = {\rm const.}
\label{Asp-b}
\ee
for all streaks, $i$.  This is in contrast to work in refs.\cite{M2001,M2002}. 
\vspace{.2cm}

In the local rest frame of the $i$-th peripheral streak, RFS$_i$, the 
streak momentum vanishes, while in the collider c.m. frame it is
$P_{iz}$ given by the pre-collision parameters,
and thus the streak rapidity in the collider c.m. frame is given by eq. (\ref{eta})
\be
<\eta_i> = 
\Artanh \frac{P_{iz}}{E_i}\ .
\ee

Based on eqs. (\ref{E_eta}) and (\ref{Asp-b})
\be
\frac{1}{2}\Delta \eta_i = 
\Arsinh \left( \frac{E_i}{2 \tau_0 e_c(\tau_0) A \cosh(<\eta_i>)}
\right) \ .
\label{delta_eta}
\ee

Knowing $<\eta_i>$ and $\Delta \eta_i$, means that 
$\eta_i$ varies between $\eta_{min}^i= <\eta_i> - \frac{\Delta\eta_i}{2}$  and 
$\eta_{max}^i = <\eta_i> + \frac{\Delta\eta_i}{2}$,
and the end points of streak $i$ on the P side will  be
\ba
z_{max}^{i,P} & = & z_{max}^c \,, \nonumber \\
t_{max}^{i,P} & = & t_{i0} + \tau_0 \cosh \eta_{max}^{i}  \,,
\label{zt_max_P}
\ea
and
\ba
z_{min}^{i,P} & = & z_{i0} + \tau_0 \sinh \eta_{min}^{i}\,, \nonumber \\
t_{min}^{i,P} & = & t_{i0} + \tau_0 \cosh \eta_{min}^{i}  \,,
\label{zt_min_P}
\ea
where $z_{i0}$ and $t_{i0}$ are still unknown.

Similarly, we can perform the calculation for the T side streaks: 
\ba
z_{max}^{i,T} & = & z_{i0} + \tau_0 \sinh \eta_{max}^{i}\,, \nonumber \\
t_{max}^{i,T} & = & t_{i0} + \tau_0 \cosh \eta_{max}^{i}  \,,
\label{zt_max_T}
\ea
and
\ba
z_{min}^{i,T} & = & z_{min}^{c}\,, \nonumber \\
t_{min}^{i,T} & = & t_{i0} + \tau_0 \cosh \eta_{min}^{i}  \,.
\label{zt_min_T}
\ea

Now the baryon density at $\tau_i=\tau_0$ can be found from 
eq. (\ref{N_eta}):
\be
n_i(\tau_0)=\frac{N_i}{\tau_0 A \Delta \eta_i} \,.
\label{n_tau0}
\ee 

The previous description of assumption {\bf (a)} was applicable for streaks 
in the reaction plane, {\it i.e.}, for $y_i=0$ coordinate. For each $y_i \ne 0$ layer
of streaks we define a new central streak with $P_{iz}^{c-y}=0$ and $<\eta_{i}^{c-y}>=0$. Then following the assumption {\bf (b)}, eq. (\ref{Asp-b}), 
we can find the $\eta$ extension of this streak, similarly to eq. (\ref{delta_eta}),
\be
\frac{1}{2}\Delta \eta_i^{c-y} = 
\Arsinh \left( \frac{E_i^{c-y}}{2 \tau_0 e_c(\tau_0) A )}
\right) \ .
\label{delta_eta_y}
\ee  
Then for all the non-central streaks corresponding to the same coordinate $y$ we can repeat the above mentioned steps.

\subsection{Matching the leading {\large $z$} 
and the mid {\large $t$} of the streaks}
\label{3B}

As one can see in Fig. \ref{f2} if the space-time points of the
central and side streaks match, the back end of the side streak 
(red hyperbola, starting at $\eta_{min}$) 
may be very far from the central streak in the 
space-time,  and 
the back end of the side streak may even be out of the light cone
of the central streak. Instead one can assume that the front end of the
side streak is at the same point as that of the central streak, but
the time coordinate of the mid point of the (red) 
side streak, $t_{imid_P}$, falls
on the hyperbola of the (blue) central streak, as  shown in 
 Fig. \ref{f2}. This mid-point corresponds to the mid geometrical rapidity, {\it i.e.}, $\eta_i=<\eta_i>$ defined in eq. (\ref{eta}).
\ba
t_{imid_P} &=& t_{i0} + \tau_0 \cosh <\eta_i> ,
\nonumber \\
z_{imid_P} &=& z_{i0} + \tau_0 \sinh <\eta_i> . 
\label{imid}
\ea
We now have to include the condition that the point $[t_{imid_P},\, z_{imid_P}]$ falls
on the hyperbola of the central streak:
\be
\tau_0^2 = t_{imid_P}^2 - z_{imid_P}^2\;
\label{tau0}  .
\ee
With eq. (\ref{imid})  this leads to the first connection between $t_{i0}$ and $z_{i0}$
\be
\tau_0^2 = \!\!
\left(t_{i0} {+} {\tau_0} \cosh <\eta_i> \right)^{2} 
- 
\left(z_{i0} {+} {\tau_0} \sinh <\eta_i> \right)^{2} .
\label{tauC}
\ee
To find these two unknowns we have to add one more equation, which comes from our assumption (a):
\be
z_{max}^{i,P} = z_{max}^c\, .
\ee

\begin{figure}[htb]     
\begin{center}
\resizebox{1.01\columnwidth}{!}
{\includegraphics{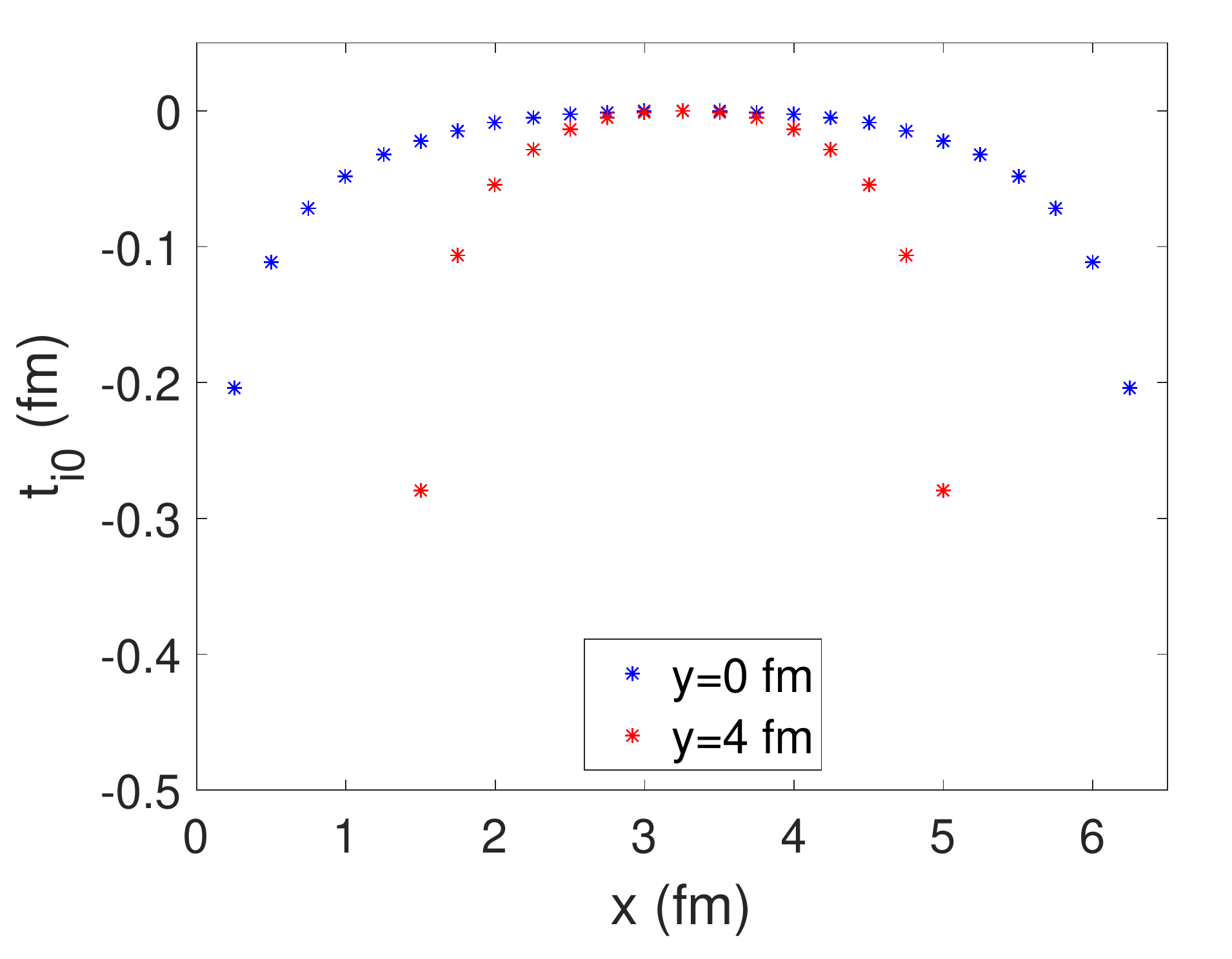}}
\caption{ (color online)
The initial time coordinates, $t_{i0}$,
of the  $i$th streaks  in the reaction 
plane $[x,z]$ for different values of $x$, for $y=0$
(blue stars) and for $y = 4$ fm (red stars).
The reaction parameters are the same as 
listed in Fig. \ref{Config-in-xz}.
}
\label{t0i}
\end{center}
\end{figure}        

Thus, for the P side we obtain:
\be
z_{i0} = z_{max}^c - \tau_0 \sinh \eta_{max}^{i}\,.
\label{zi0}
\ee
By keeping the
longitudinal, forward $z$-positions of the streak ends at the same point,
we know $z_{i0}$ from eq. (\ref{zi0}). Then inserting  $z_{i0}$ into 
eq.  (\ref{tauC}), we can get $t_{i0}$. Thus, 
{\bf we get both} $t_{i0}$ and $z_{i0}$ for each streak on the P side.

The T side can be calculated similarly by matching the back ends
($z_{min}$) of the side streaks.

Globally, for the whole collision, our assumptions lead to a rather compact IS in the space-time, as the centers of each side streak fall on the hyperbola of the central streak.

\subsection{Results for the first step of the model: streaks at $\tau_i=\tau_0$}
\label{3C}

To better illustrate how the model works we performed a calculation simulating a
Au+Au reaction at $100+100$ GeV/nucl energy and impact parameter
$b= 0.5 (R_{Au} + R_{Au}) = 6.5$ fm. The model parameters are fixed at $\tau_0=1.0$ fm  and $\Delta \eta_c = 2$, 
which leads to the energy density $e_c(\tau_0)  = 156.31$ GeV/fm$^3$.  
The  results are shown in  Figs. \ref{Config-in-xz}, \ref{Config-in-xeta}, \ref{eta-i}, and \ref{Delta-eta}.  Note that, strictly speaking, Figs. \ref{Config-in-xz}, \ref{Config-in-xeta} serve only for a qualitative 
understanding of the model, since all the streaks are plotted at the moment when their $\tau_i=\tau_0$. The method of constructing a proper model of the collision and presenting the different distributions at a given time in the Laboratory system or at one global proper time will be discussed in the next two sections.

Each of our final streaks has a scaling expansion flow in its own frame. Each streak is homogeneous and finite; all
conservation laws, including the angular momentum, are exactly satisfied by construction, at least at the moment $\tau_i=\tau_0$. 
Thus our initial state model includes local shear and vorticity.

Our IS reflects qualitatively the 
behavior of the parton kinetic models: the blue streaks of Fig. \ref{Config-in-xeta} should be compared with Fig. \ref{Streaks}.

The  proper time evolution of the energy density and baryon density of the given streak is given by the following equations:
\be
\frac{d e_i}{d \tau_i}=-\frac{e_i+P_i}{\tau_i}\,, \quad \frac{d n_i}{d \tau_i}=-\frac{n_i}{\tau_i}\,,
\label{Bjorken}
\ee   
where the pressure $P_i$ is given by the equation of state,  $P_i=e_i/3$. 
The initial conditions are given at  $\tau_i=\tau_0$, $e_i(\tau_0)$, $n_i(\tau_0)$ by eqs. (\ref{Asp-b},\ref{n_tau0}). This system can be solved easily:
\be
e_i(\tau_i)=e_i(\tau_0)\left(\frac{\tau_0}{\tau_i}\right)^{4/3}\,, \quad
n_i(\tau_i)=n_i(\tau_0)\left(\frac{\tau_0}{\tau_i}\right)\,.
\label{bjor-sol}
\ee

It is important to remember that if we want to have a finite volume fireball, we need to put some boundaries on the system. Here we assume that our system, {\it i.e.}, given final streak $i$,  described by the Bjorken model, is situated in the spacial domain $\eta^i_{min}<\eta_i<\eta^i_{max}$. Within these boundaries the system is uniform along $\tau_i = const$ hyperbolae due to model assumptions, while outside we have vacuum with zero energy and baryon density as well as pressure. Thus, we have a jump, a discontinuity on the boundary, which remains during all the evolution. Certainly, to prevent matter expansion through such a boundary (due to strong pressure gradient) some work is done on the boundary surface \cite{Bjorken_work}. One can think about it as putting some pressure to the surface with the vacuum, exactly the one which would remove discontinuity, then work is done by the expanding system against this pressure. 

This actually means that although at the moment $\tau_i=\tau_0$  the energy density is taken in such a way that the energy and momentum conservation laws are satisfied, at any other moment of the proper time the energy is not explicitly conserved, because of the fixed $\eta$ boundaries: some is lost (for $\tau_i > \tau_0$), or it is also possible that some is  gained  (for $\tau_i < \tau_0$). Thus, for the overall collision IS, which we will be presented in the following sections at a given time in the Laboratory system or at one global proper time, which will require to some over different local $\tau_i$s, bigger or smaller than $\tau_0$, the total energy is strictly speaking not conserved. Although the difference is not that big for the IS parameterizations presented in the next section, since we aware of this problem and trying to control it: the conservation laws are satisfied up to 3$\%$ of accuracy (usually better), which incudes also the errors coming from numerical gridding.

\begin{figure}[htb]     
\begin{center}
\resizebox{1.01\columnwidth}{!}
{\includegraphics{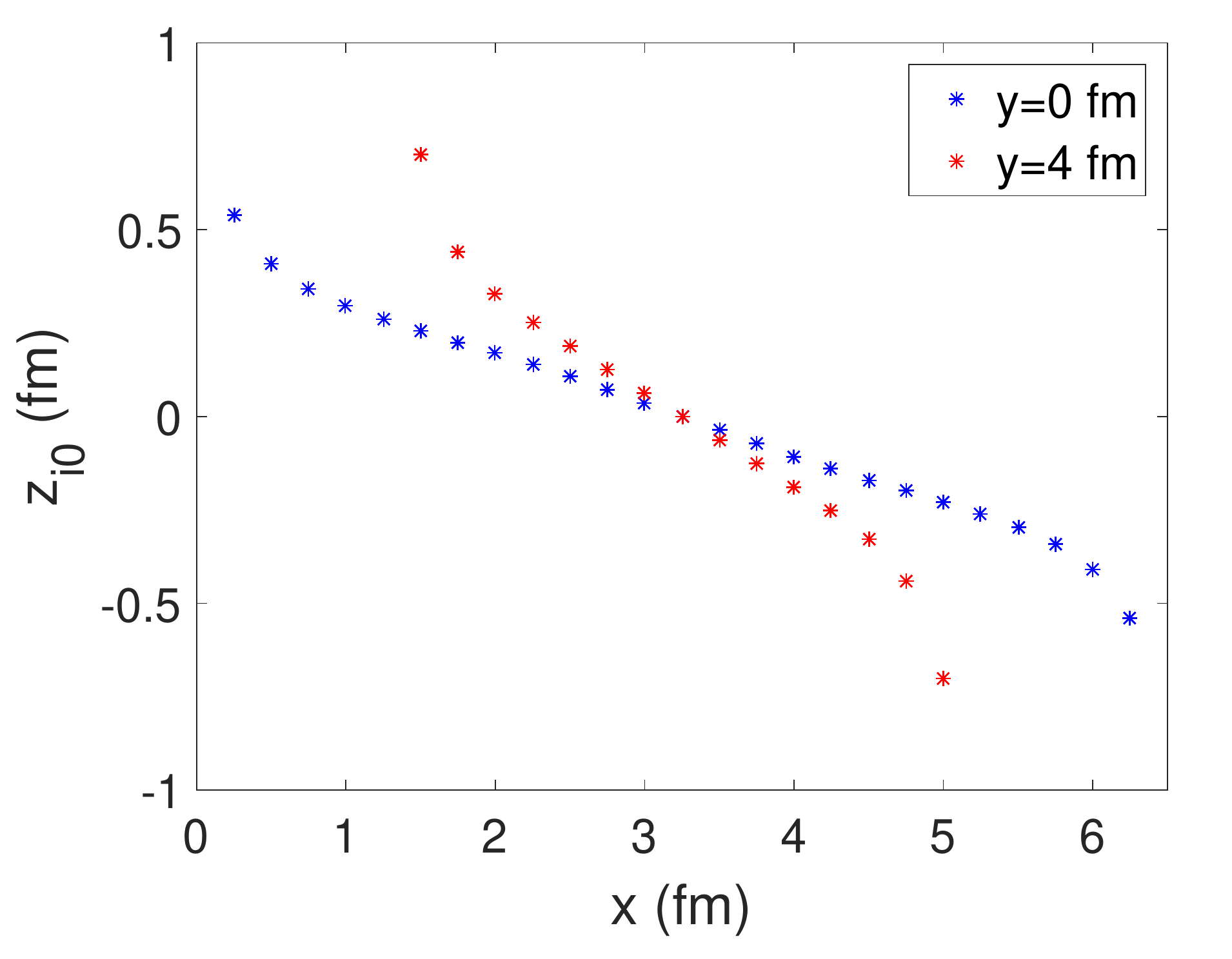}}
\caption{ (color online)
The initial $z_{i0}$ coordinates of the $i$th streaks in the reaction 
plane $[x,z]$ for different values of $x$, for $y=0$
(blue stars) and for $y = 4$ fm (red stars).
The reaction parameters are the same as 
listed in Fig. \ref{Config-in-xz}.
}
\label{z0i}
\end{center}
\end{figure}        

\section{Implementation in Cartesian coordinates}

In our model in the first stage each of the streaks should be treated separately with the Bjorken model with Milne coordinates, [$\tau, x,y,\eta$];  
the initial state definition lies on different $\tau_i = \tau_0$ hyperbolae. 
In order to present the global initial state for the whole collision the local baryon and energy densities along these streaks as well as the
local flow velocities should be taken at some global initial state hypersurface in space time.

In principle the initial state can be defined on any time-like hypersurface,
{\it i.e.}, with any hypersurface with time-like normal vectors.  
In the general case this can be a complex curved time-like hypersurface; however the PICR code, for example, allows the 
implementation of the curved IS hypersurface. 

If the fluid dynamical code cannot handle a complex curved (in general case) time-like hypersurface 
and if the IS model and the FD model have different EoS, or if the IS model has no local equilibrium and therefore
has no EoS,
one has to use the matching conditions between the initial state
reference frame and the fluid dynamical model's reference frame,
as described in ref. \cite{Cheng2010}.

In this particular work we would like only to illustrate qualitatively the proposed new model for the IS, and therefore we will stay with the simplest choices of a time-like hypersurface for the transition 
from the initial state model to the fluid dynamics, which are $t=t_{IS}=const.$ or  $\tau=\tau_{IS}=const.$

The use of  Milne  coordinates $(\tau, x, y,\eta)$ in hydrodynamical calculations  requires additional work that is outside the scope of this paper.
The  effects of an increasing cell size 
in the longitudinal direction during the calculation which leads to 
increasing numerical viscosity and dissipation as well as anisotropic
viscosity are not understood.  Analysis of these effects of a changing grid size on dissipation at relativistic energies is lacking and is much needed.
Thus, using Cartesian coordinates with a constant and isotropic grid
is advantageous for avoiding numerical anisotropy and other artifacts.

\begin{figure}[htb]     
\begin{center}
\resizebox{1.01\columnwidth}{!}
{\includegraphics{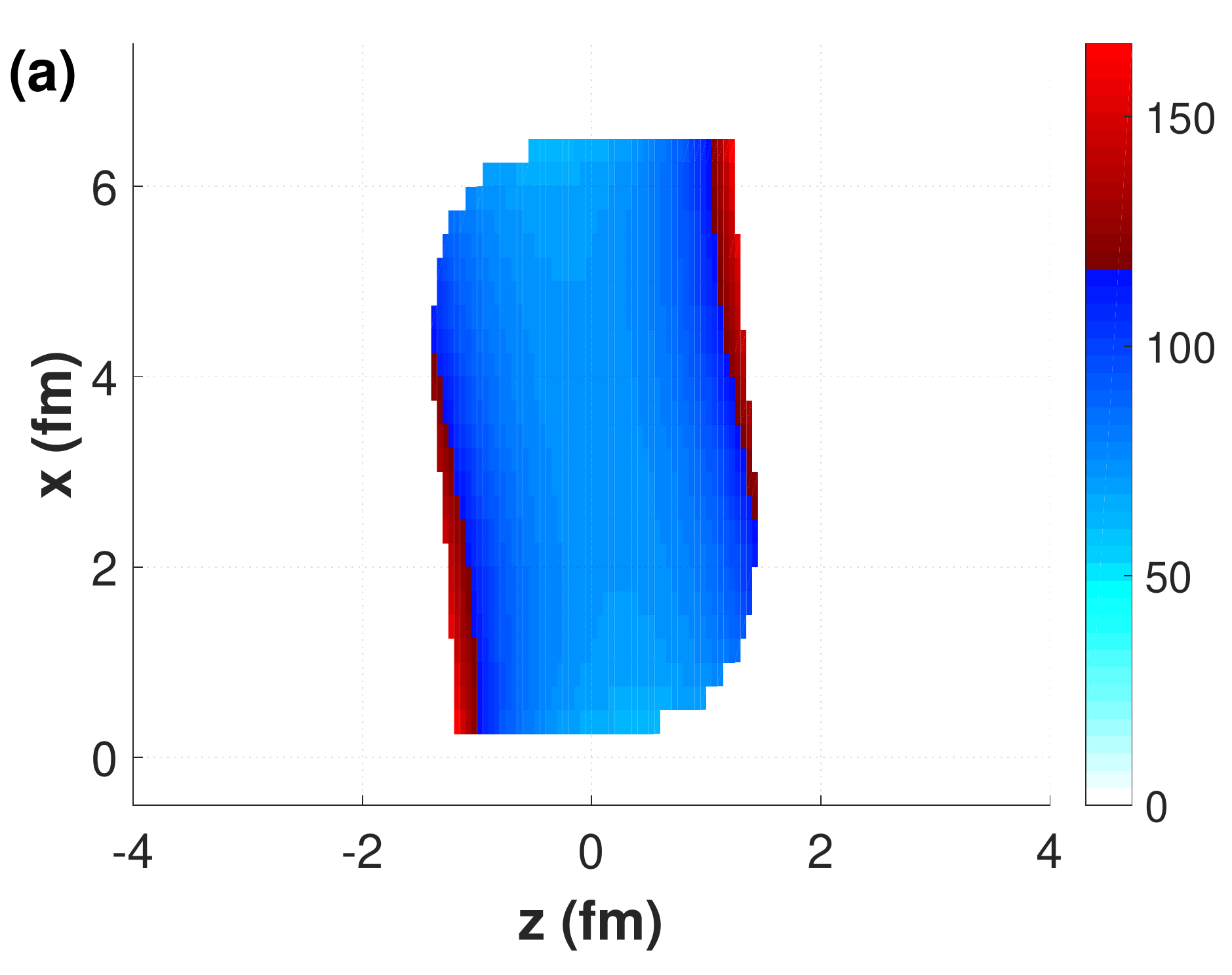}}
\resizebox{1.01\columnwidth}{!}
{\includegraphics{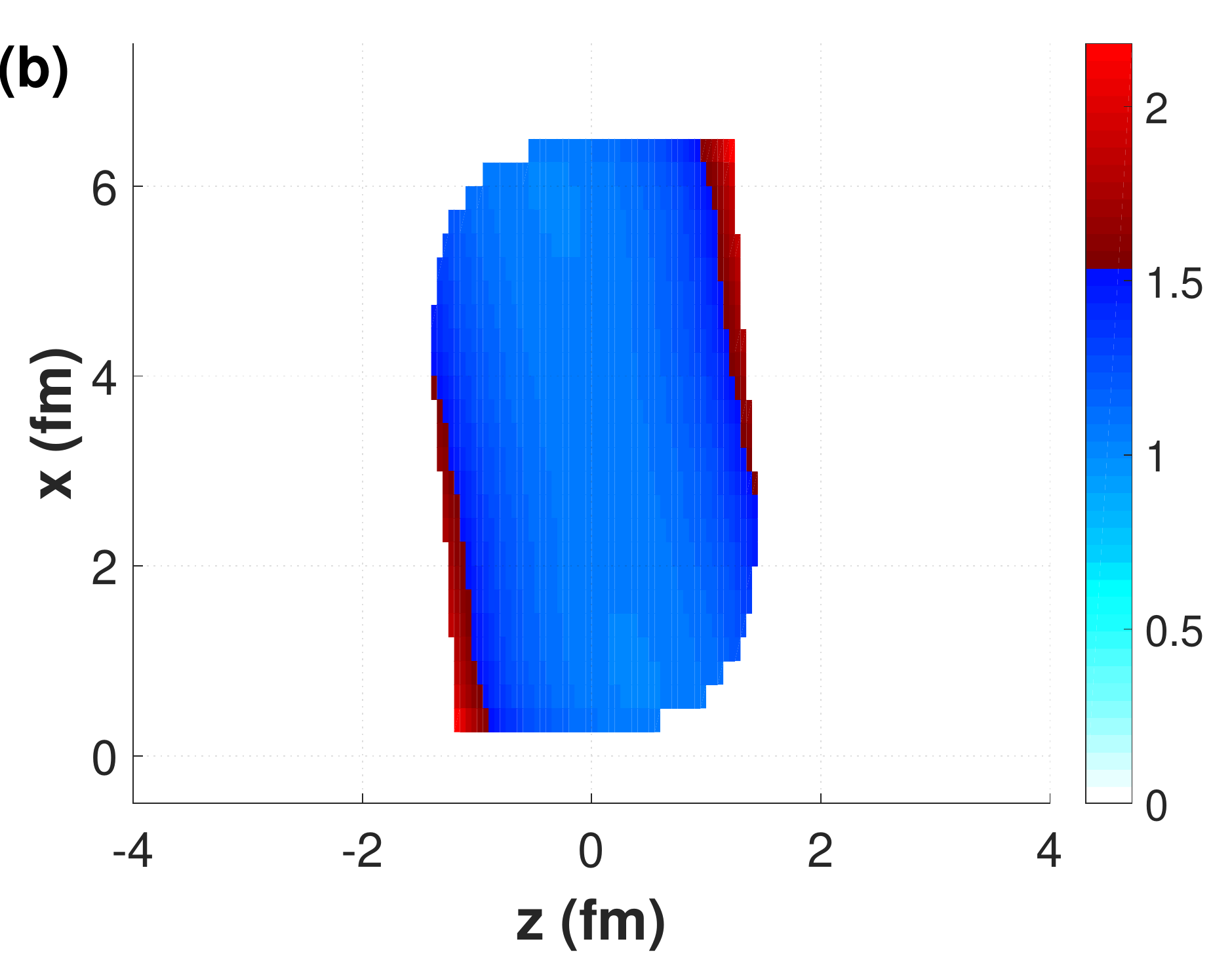}}
\caption{ (color online)
The reaction plane, $y=0$, $[x,z]$  plot of the energy density 
(upper panel) and baryon density (lower panel) in 
units of GeV/fm$^3$ and 1/fm$^3$ correspondingly,
propagated to the constant time, $t=t_{IS} = 1.78$ fm/c 
hypersurface. The resulting density distributions shows a maxima at the both the forward and backward edges of the
reaction plane.
Although the energy density and baryon density are uniform
at $\tau_i = \tau_0$ for each streak in its own frame as shown in
Fig. \ref{Config-in-xz}, the observed space-time dependence arises from 
the propagation to the $t=t_{IS}$ hypersurface.
The reaction parameters are the same as 
listed in Fig. \ref{Config-in-xz}.
}
\label{e-XZ}
\end{center}
\end{figure}        

 For a fluid dynamics (FD) model in Cartesian coordinates $(t, x, y, z)$ 
there is an obvious choice of using the Lab or collider c.m. reference frame.   We can define the transition surface between the IS model and the FD
model in the Cartesian coordinates and can propagate the IS model 
solution up to this transition hypersurface. This will result in an
initial state where the space-time points of the transition hypersurface
do not have a constant $\tau_i$ from the origin of the $i$-th peripheral
streak. Furthermore, each different peripheral streak, $i$, will have a
different space-time origin.

Let us choose a constant time, $t=t_{IS}=const.$, for the 
initial state hypersurface. Then we propagate (or cut) the
initial state model up to this hypersurface from the initial $\tau_i=\tau_0$
hyperbolae. For a given $z_{IS}$ longitudinal coordinate of this 
$t=t_{IS}$ hypersurface for the $i$-th streak the proper time
from its origin, $[z_{i0}, t_{i0}]$ will be 
\be
\tau_i '(z_{IS}) = \sqrt{(t_{IS}- t_{i0})^2 - (z_{IS}- z_{i0})^2\,,}
\label{tau_prime}
\ee
where
\be
z_{IS} = z_{i0} + (t_{IS} - t_{i0}) \tanh  \eta_i \,,
\ee
and $\eta_i$ can vary in the interval $[\eta_{min}^{i,T},\eta_{max}^{i,P}]$.

Now using  Bjorken hydrodynamic solution one can get 
the invariant scalar energy and baryon densities on this
hyperbola:
\be
e_{IS}(\tau_i ') = e_c(\tau_0) \left( \frac{\tau_0} {\tau_i '}\right)^{\!\!4/3}
{\rm and} \ \
n_{IS}(\tau_i ') = n_i(\tau_0) \left( \frac{\tau_0} {\tau_i '}\right)\,.
\label{bj_sol_prime}
\ee

\begin{figure}[htb]     
\begin{center}
\resizebox{1.01\columnwidth}{!}
{\includegraphics{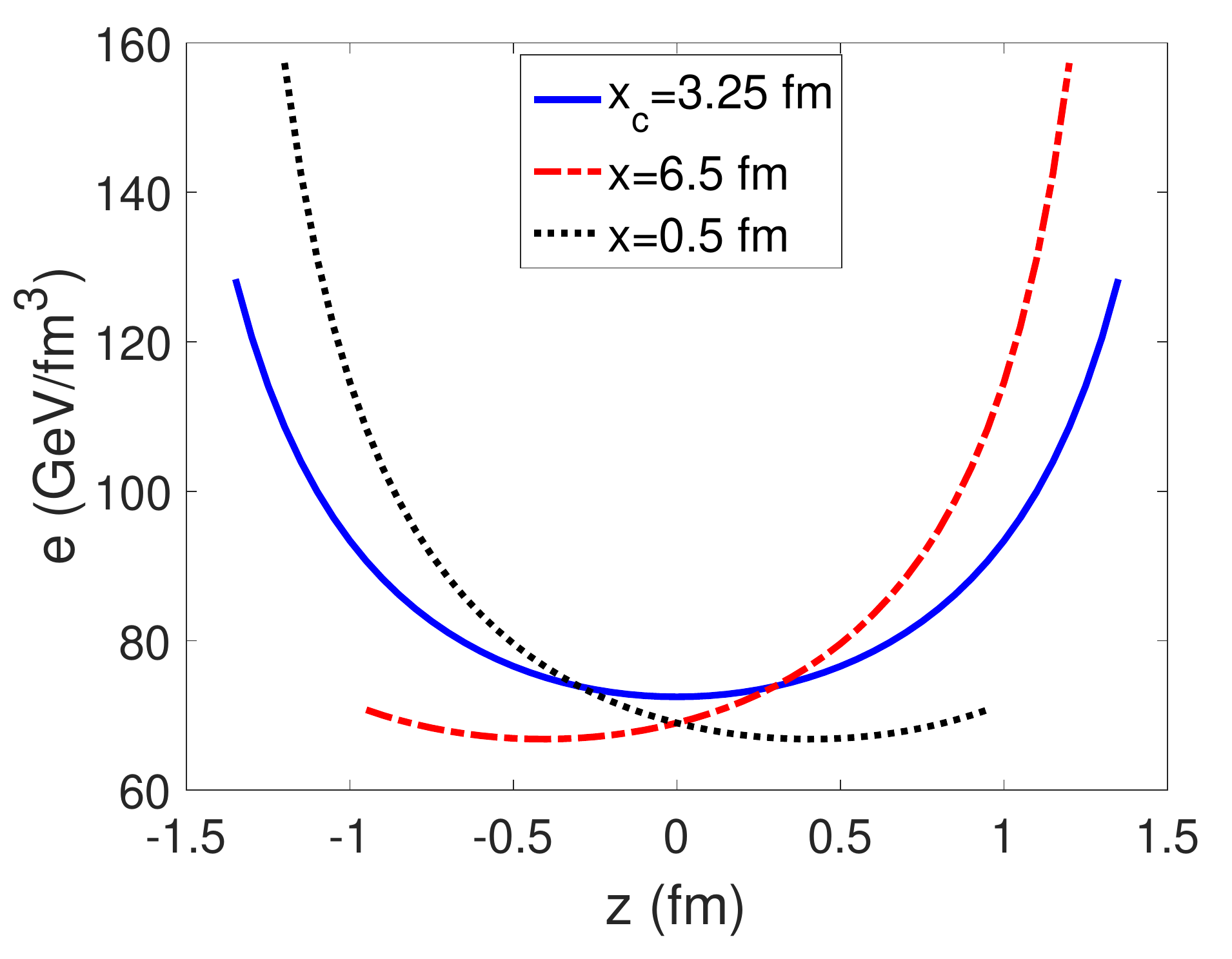}}
\caption{ (color online)
The energy density distribution along the $z$ direction, for the 
central streak at $x = 3.25$ fm and for the peripheral streaks at
$x = 0.5 \, \& \, 6.5$ fm.
The matter of fluid elements was propagated to the constant time, 
$t_{IS} = 1.78$ fm/c, hypersurface. 
This example is calculated for the same reaction and parameters 
that are listed in Fig. \ref{Config-in-xz}.
}
\label{finez}
\end{center}
\end{figure}        

\begin{figure}[htb]     
\begin{center}
\resizebox{1.01\columnwidth}{!}
{\includegraphics{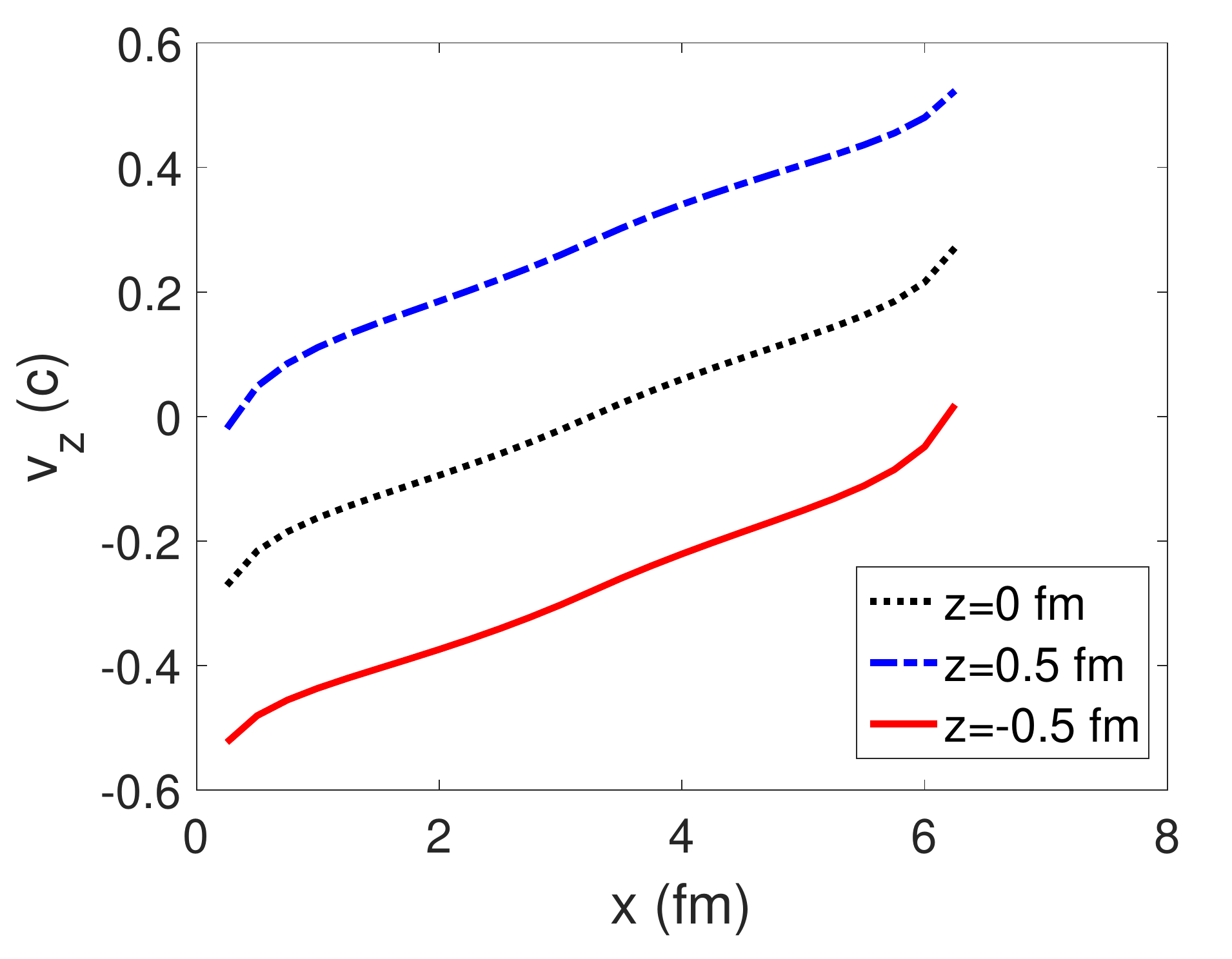}}
\caption{ (color online)
The $z$-directed velocity distribution versus the $x$ position 
(at $y = 0$), in the $z = 0$ fm
central plane (dash line)
propagated to the constant time, $t_{IS} = 1.78$ fm/c
hypersurface. 
The velocity distribution for the 
$z = \pm 0.5$ fm forward/backward shifted positions 
are shown by dash-dotted and solid lines respectively.
The reaction parameters are the same as listed in Fig. \ref{Config-in-xz}.
}
\label{finvx}
\end{center}
\end{figure}        

We perform a simulation of the 
Au+Au reaction at $100+100$ GeV/nucl energy and impact parameter
$b= 0.5 (R_{Pb} + R_{Pb}) = 6.5$ fm, as shown in Fig. \ref{Config-in-xz}.  The model parameters are $\tau_0=1.0$ fm  and $\Delta \eta_c = 2$.
We end our simulation of the IS for this reaction at a constant time hypersurface, $t_{IS} = 1.78$ fm. This is actually a minimal possible time for such a calculation, namely $t_{IS}=$Max$\{t_{max}^i\}$. Choosing any smaller $t_{IS}$ would lead to the situation  when some of $\tau_i '$s, calculated according to eq. (\ref{tau_prime}), would be smaller then the Bjorken initial state time $\tau_0$. 

Our model with the above choice of parameters leads to a compact IS, substantially different from the IS of Ref. \cite{M2001, M2002}: one can compare, for example the energy density distribution of Fig. \ref{e-XZ} with Fig. 13(A) of Ref. \cite{M2002}.    As one can see in Figs. \ref{e-XZ} and \ref{finez} the energy density in the middle region will be less due to the propagation to the corresponding larger proper time, $\tau_i '$.

\begin{figure}[htb]     
\begin{center}
\resizebox{1.01\columnwidth}{!}
{\includegraphics{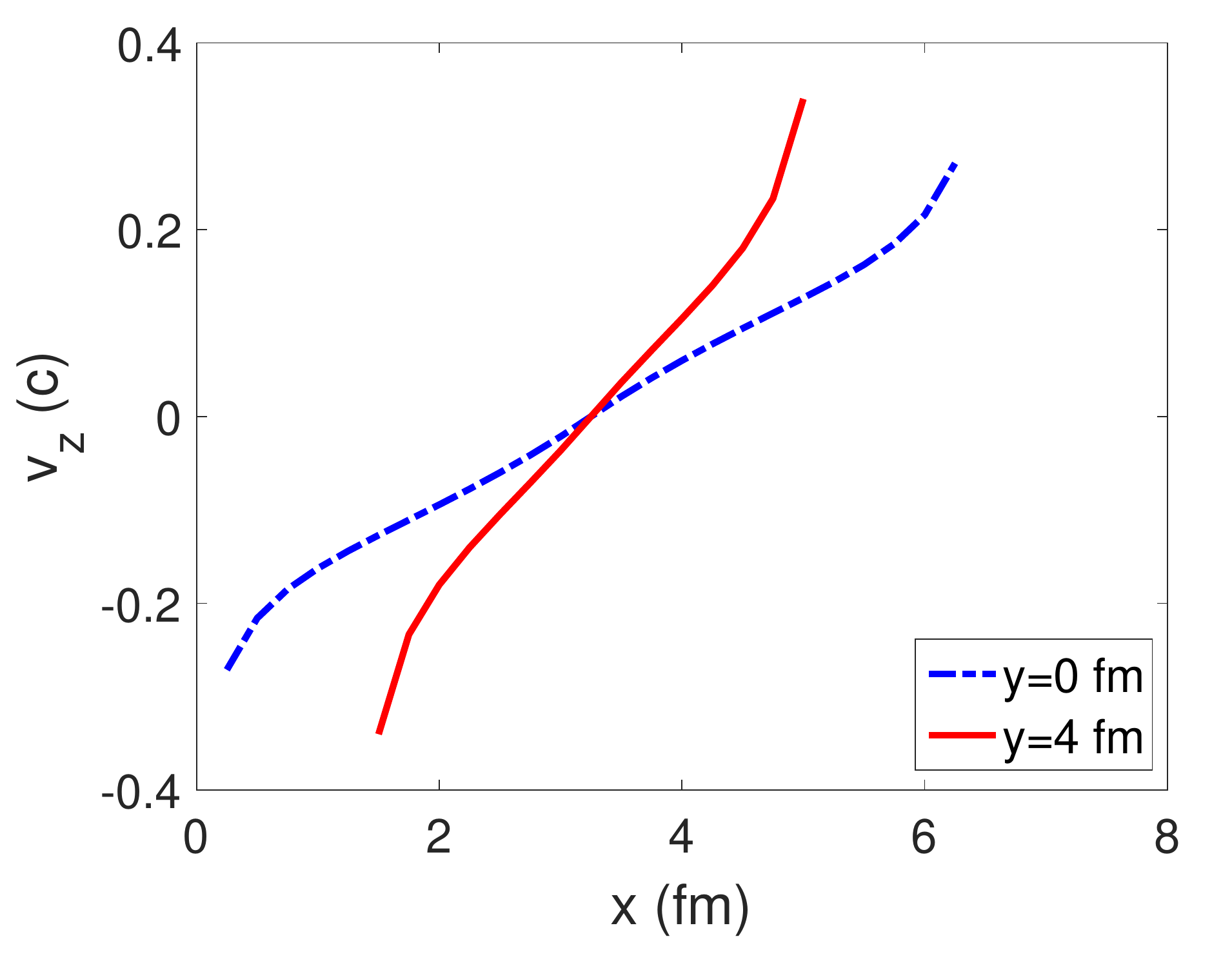}}
\caption{ (color online)
The velocity, $v_z$, distribution in $x$ direction (at $z = 0$), 
for the central layer, $y = 0$ fm (dash-dot line), and for the 
 peripheral layer at $y = 4$ fm (solid line),
propagated to the constant time, $t_{IS} = 1.78$ fm/c hypersurface. 
The reaction parameters are the same as 
listed in Fig. \ref{Config-in-xz}. Interestingly the longitudinal
shear among the neighboring peripheral $x$ layers is much bigger than in the center.
}
\label{vxdy}
\end{center}
\end{figure}        

The structure of the net baryon density distribution is very similar to the energy density distribution, 
as shown in Fig. \ref{e-XZ}. The propagated net baryon density shows maxima at the forward and 
backward edges of the matter in the reaction plane.  The maximum value of the nucleon number is $n \approx 2.7$ fm$^{-3}$.

As one can see in Fig. \ref{finvx} the present model shows considerable shear in the velocity field.
In the center of the reaction plane, in the direction of the impact parameter
vector, $x$, the upper (positive) side shows a forward motion  (positive velocities $v$) while the lower side shows negative velocities. Further forward in the beam  direction, (at $z=0.5$ fm) the velocity profile is identical but shifted in velocity to higher  positive values due to the longitudinal Bjorken expansion on the model.  On the opposite side (at $z=-0.5$ fm) the shift is opposite due to the Bjorken expansion.

The side layers that are parallel to the reaction plane
at finite $y$ values, show the same shear flow profile,
but with higher shear, $\partial v_z / \partial x$, see 
Fig. \ref{vxdy}.

\begin{figure}[htb]     
\begin{center}
\resizebox{1.01\columnwidth}{!}
{\includegraphics{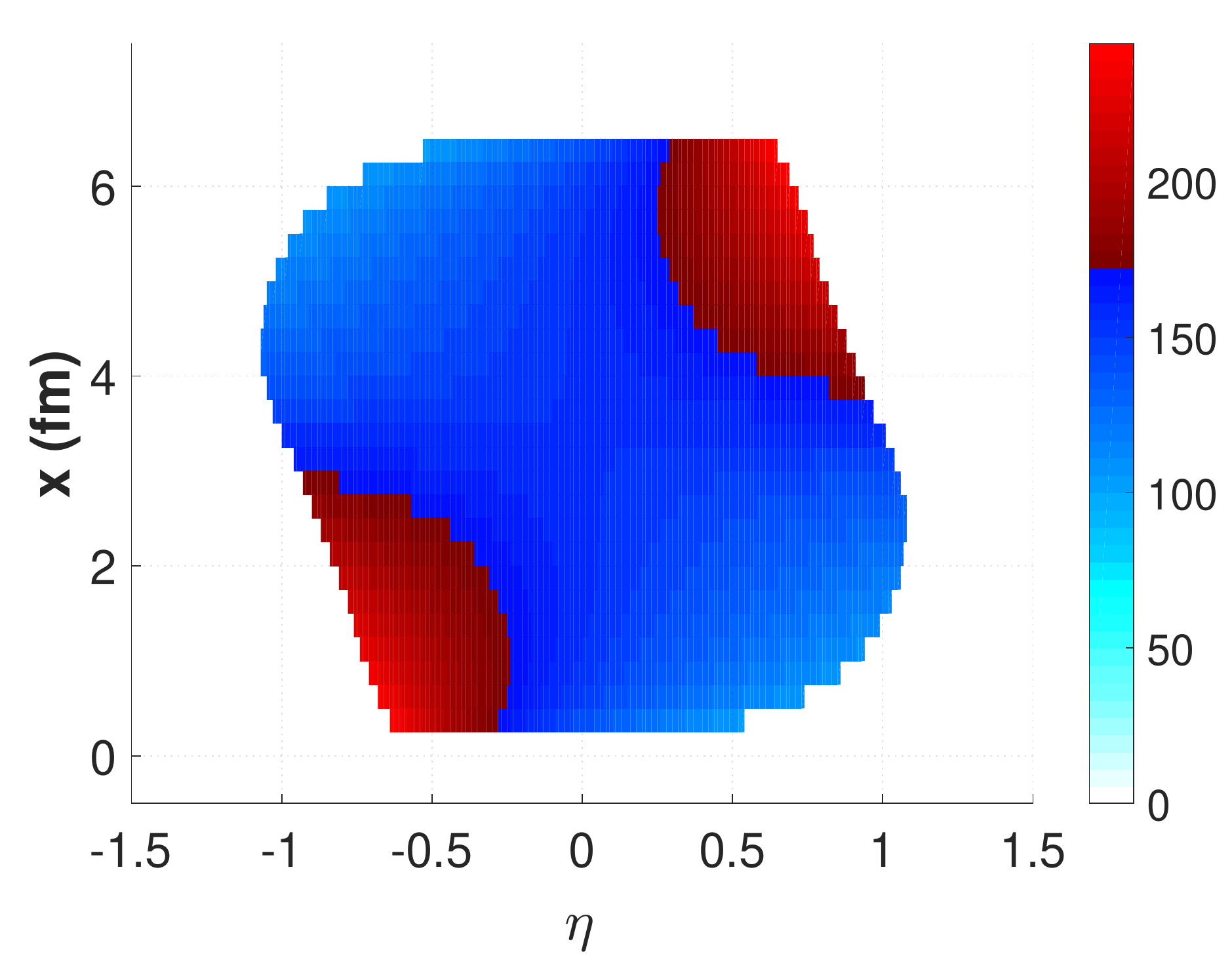}}
\caption{ (color online)
The reaction plane $y=0$, $[x,z]$  plot of the energy density,
$e(x,\eta)$, in units of GeV/fm$^3$,
propagated to the constant proper time, $\tau_c=\tau_0 = 1.0$ fm/c 
hypersurface. The $(\tau_i,\eta_i)$ points of each particular streak are propagated forward and backward to the overall  $\tau_{c} = 1.0$ fm/c hyperbola, as explained in the text.
The propagated initial density shows maximal
energy densities of $e \approx 240$ GeV/fm$^3$ at the 
forward and backward edges of the reaction plane, although $e_c(\tau_0) = 156.31$ GeV/fm$^3$.
The reaction parameters are the same as 
listed in Fig. \ref{Config-in-xz}.
}
\label{Tauf-e-ex}
\end{center}
\end{figure}        

\section{Implementation in Milne coordinates
{\Large $\tau$}, {\large $x, y$}, {\Large $\eta$} }\label{Milne}

We now show how our IS can be implemented in a fluid dynamical model in
$x,y,\eta,\tau$ coordinates.  
In these coordinates the numerical solution can be performed
in the frame of the central streak. In other words our model will give an initial state for further FD evolution on the  $\tau_c=\tau_{IS}=const.$ hypersurface.

Using the c.m. $x,y,\eta_c,\tau_c$ coordinates one can calculate for each
space-time point of the hypersurface the corresponding $x,y,z,t$ coordinates using eqs. (\ref{tz1}), and then relating those with any $x,y,\eta_i,\tau_i$ RFS$_i$ frame.

\begin{figure}[htb]     
\begin{center}
\resizebox{1.01\columnwidth}{!}
{\includegraphics{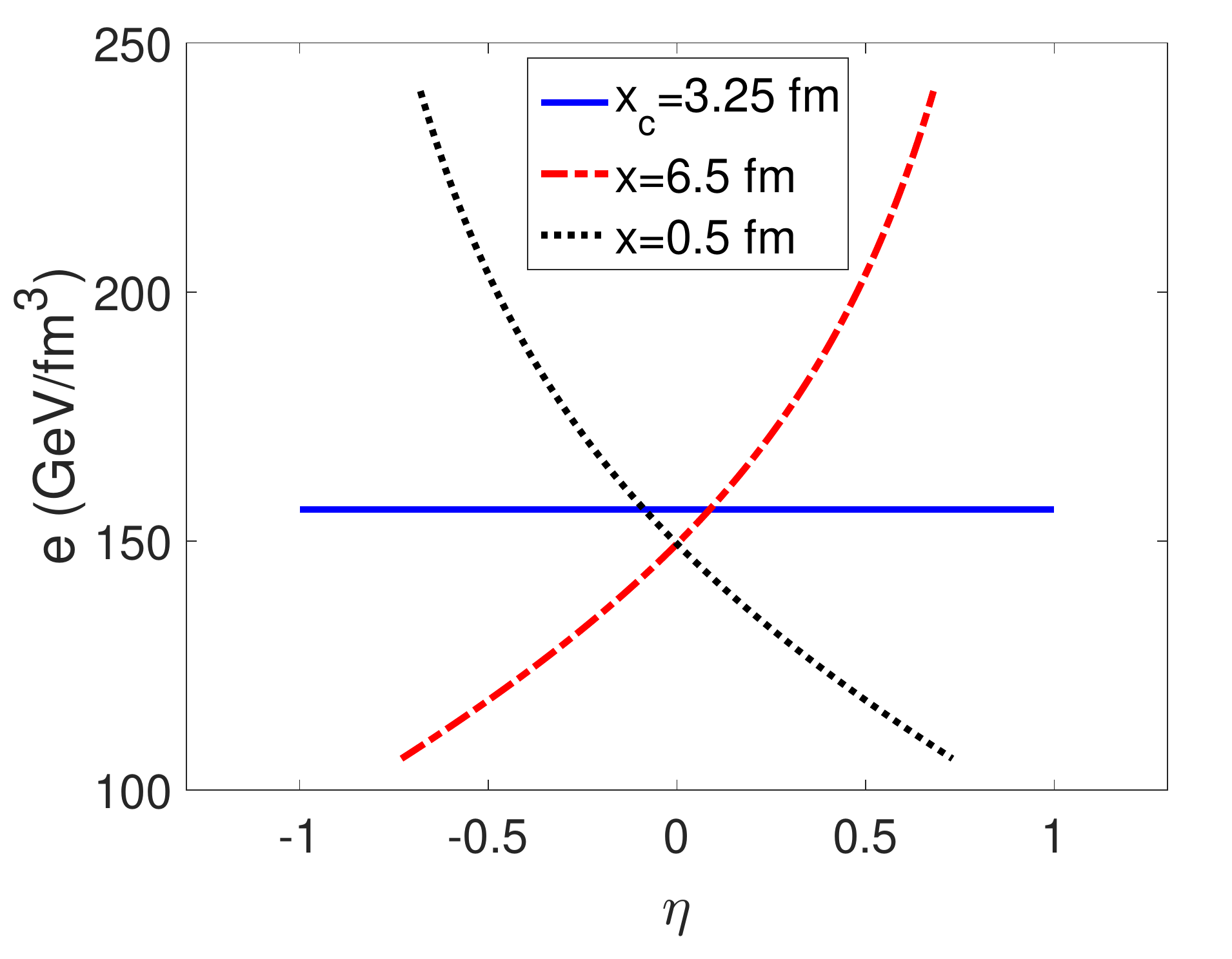}}
\caption{ (color online)
The energy density, $e$, distribution versus $\eta$,
for the central streak at $x_c = 3.25$ fm and for 
peripheral streaks at  $x = 6.5$ and $0.5$ fm,
propagated to the constant proper time, $\tau_{c} = 1.0$ fm/c 
hypersurface. The propagated initial density shows maximal and minimal
energy densities at the forward and backward edges of the
peripheral streaks.
The reaction parameters are the same as 
listed in Fig. \ref{Config-in-xz}.
}
\label{Tauf-e-e}
\end{center}
\end{figure}        

\begin{figure}[htb]     
\begin{center}
\resizebox{1.01\columnwidth}{!}
{\includegraphics{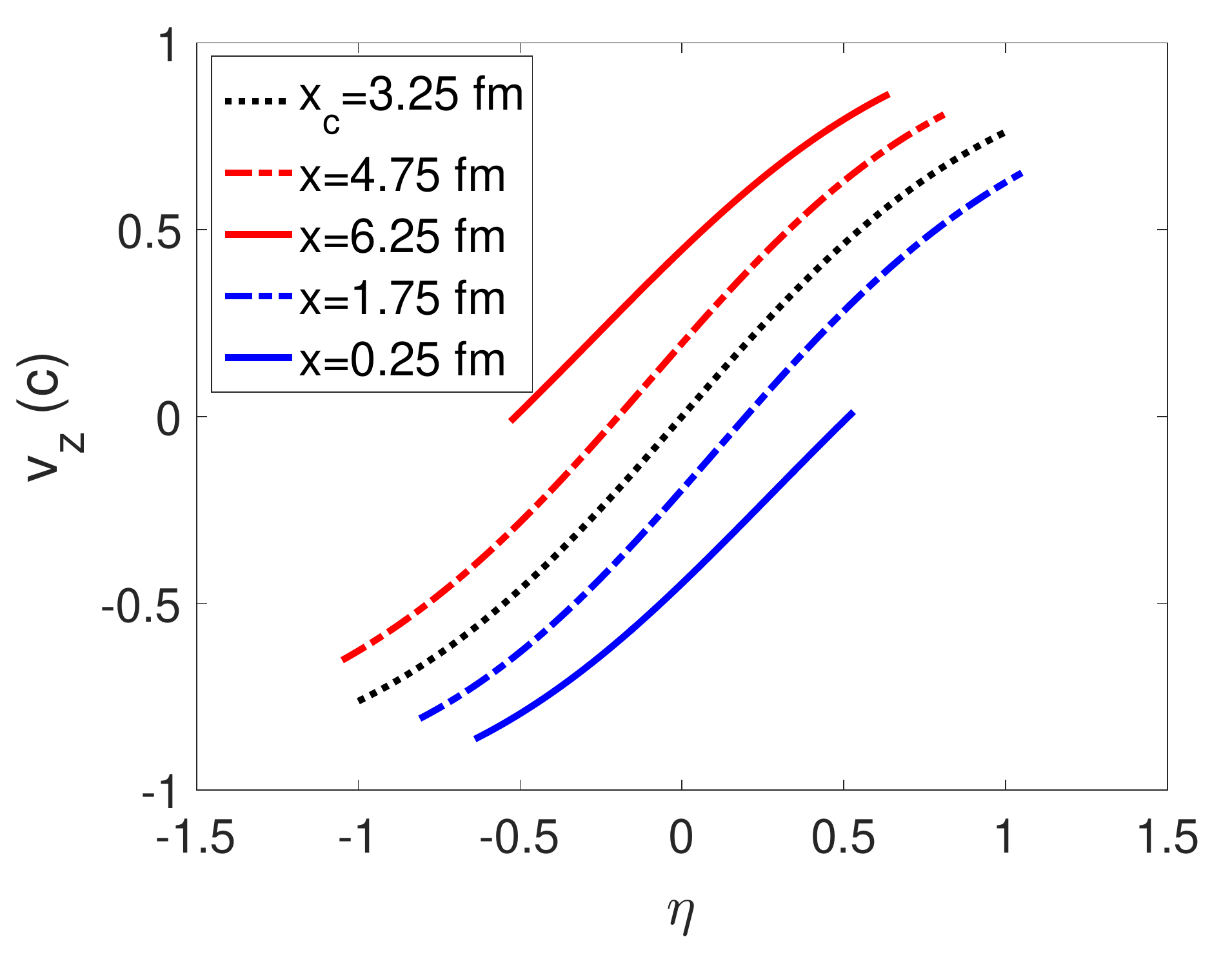}}
\caption{ (color online)
The $z$-directed velocity distribution versus the $\eta$ coordinate 
(at $y = 0$),
for the central streak (at $x_c = 3.25$ fm). Semi-peripheral streaks 
($x = 1.75 \, \& \, 4.75$ fm) 
and peripheral streaks ($x = 0.25 \, \& \,6.25$ fm), 
propagated to the constant proper time, $\tau_{c} = 1.0$ fm/c 
hypersurface are also shown.
The reaction parameters are the same as 
listed in Fig. \ref{Config-in-xz}.
}
\label{Tauf-v-e}
\end{center}
\end{figure}        

In the central streak frame for any point on a $\tau_c = \tau_{IS}$ hyperbola we have ($t_{c0}=0$, $z_{c0}=0$)
\be
t_{IS} = \tau_{IS}\cosh \eta_c\,, \quad 
z_{IS}= \tau_{IS} \sinh \eta_c\,.
\label{deftz}
\ee
In the frame of the $i$-th side streak on the other hand
$$
\tau_i ' = \sqrt{\!\left( t_{IS} - t_{i0} \right)^{2}\! -      
      \!\left( z_{IS} - z_{i0} \right)^{2} } =
$$
\be
    \sqrt{\!\left( \tau_{IS} \cosh \eta_c - t_{i0} \right)^{2}\! -      
      \!\left( \tau_{IS} \sinh \eta_c - z_{i0} \right)^{2} }\! ,
\label{eq43}
\ee
and the corresponding rapidity in the $i$-th peripheral
frame is
\ba
\eta_i ' &=& 
      \Artanh\left(\frac{\tau_{IS} \sinh \eta_c - z_{i0}}{\tau_{IS} \cosh \eta_c - t_{i0}}	
\right) \ .
\label{eq44}
\ea
We know that for each streak $i$, the geometrical rapidity should be within 
the limits $[\eta_{min}^i,\eta_{max}^i]$. 
Imposing these conditions on the $\eta_i '$, given by eq. (\ref{eq44}), one can find the corresponding limits for the $\eta_c$ for the $i$-th streak. 
\be
\eta_c  \leq \eta_{max}^{i} + \Arsinh \left(\frac
{z_{i0} \cosh \eta_{max}^{i} - t_{i0} \sinh \eta_{max}^{i}}{\tau_{IS}}
\right) \,,
\label{eq48}
\ee
and
\be
\eta_c  \geq \eta_{min}^{i} + \Arsinh \left(\frac
{z_{i0} \cosh \eta_{min}^{i}  - t_{i0} \sinh \eta_{min}^{i}}{\tau_{IS}}
\right) \,.
\label{eq49}
\ee
The detailed discussion on propagation and rapidity limits can be seen 
in the Appendix.

Now with $\tau_i '$, given by eq. (\ref{eq43}), we can calculate in the IS model
energy and baryon densities
at the pre-transition
side of the IS/FD transition hypersurface, see eqs. (\ref{bj_sol_prime}).

\begin{figure}[htb]     
\begin{center}
\resizebox{1.01\columnwidth}{!}
{\includegraphics{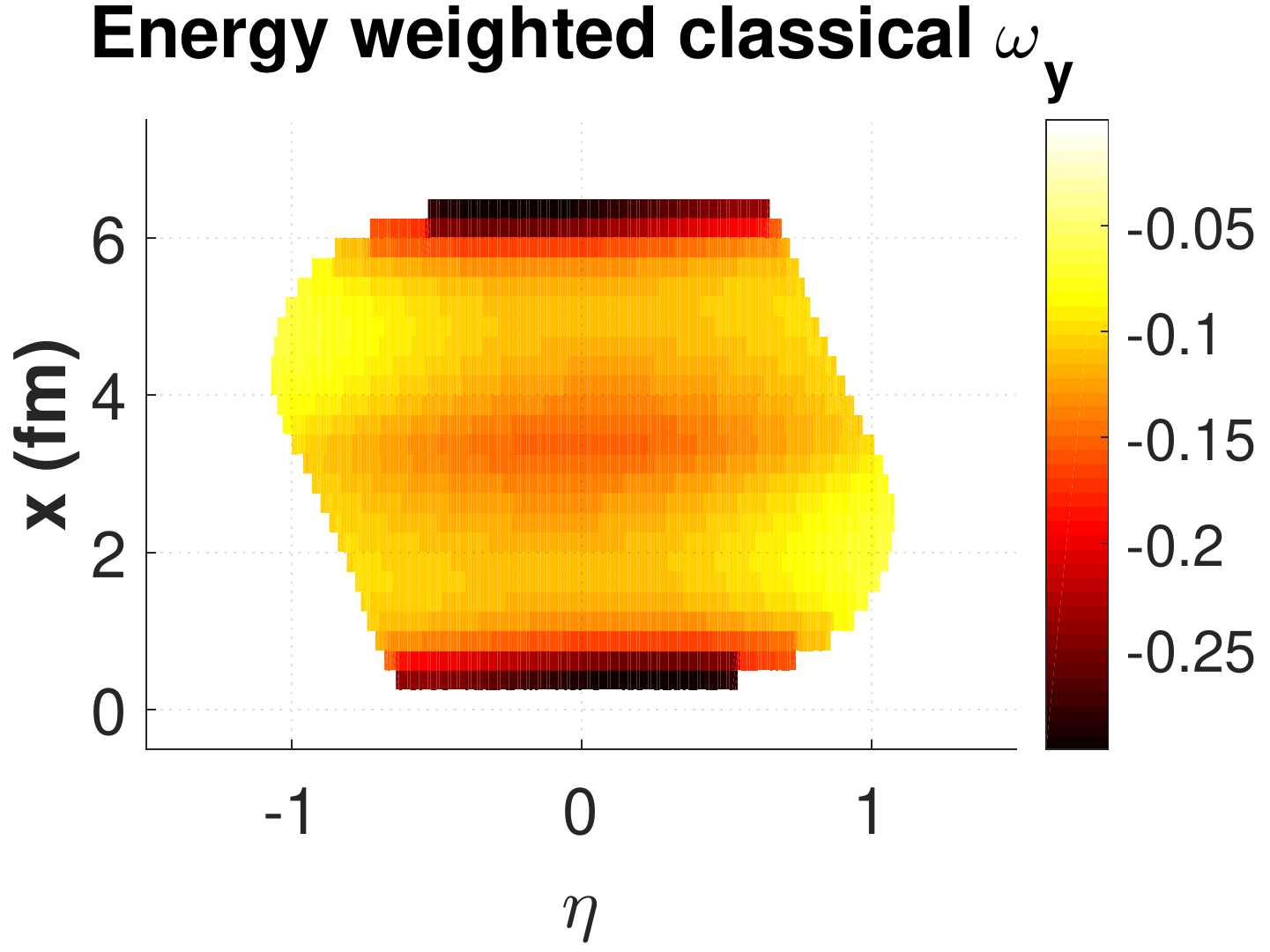}}
\caption{ (color online)
The $y$-component of the classical energy weighted vorticity
on the $[x,\eta]$ plane, at $y=0$. This is the dominant component of vorticity,
and it points everywhere in the $-y$ direction. This example is
obtained by propagation of the initial configuration to the
 hypersurface at $\tau_c = \tau_0 = 1.0$ fm/c. The at the upper
and lower edges the vorticity approaches -0.3.
}
\label{vort-ew_y}
\end{center}
\end{figure}        

The quantity $\tau_{IS}$ is a parameter of our model. For illustrative purposes of this particular study we have chosen $\tau_{IS}=\tau_0$.

The resulting energy density distribution is shown in Figs. \ref{Tauf-e-ex} and \ref{Tauf-e-e}.
For the central streak it is now flat and equal to $e_c(\tau_0) = 156.31$ GeV/fm$^3$. The non-central streaks show a strong asymmetry peaking forward or backward. 
The propagated initial density shows maximal
energy densities of $e \approx 240$ GeV/fm$^3$ at the 
forward and backward edges of the reaction plane. Such high densities, 
higher than the energy density at $\tau_i=\tau_0$ ($e_i(\tau_0) = 156.31$ GeV/fm$^3$), are reached, because for 
the $\tau_{IS}=\tau_0$ IS hypersurface the $\tau_i '$s, calculated according to eq. (\ref{eq43}), can be smaller than the Bjorken initial state time $\tau_0$, and thus $e_{IS}(\tau_i ') = e_c(\tau_0) \left( \frac{\tau_0} {\tau_i '}\right)^{\!\!4/3} > e_c(\tau_0)$. For the above formulae it is assumed that the Bjorken solution is formally valid even in the pre-equilibrium stages of the reaction $\tau_i '<\tau_0$. Such an assumption may be questionable, but in this work we only aim to illustrate our initial state model, and follow it for simplicity. 
 
The overall energy density distribution is a bit smoother for the 
$\tau=const.$ IS than for the $t=const.$ IS.
This can be clearly seen comparing Figs. \ref{finez} and   \ref{Tauf-e-e}.

The flow velocity distribution in the final streaks along the beam axis as a function of $\eta$ is shown in Fig. \ref{Tauf-v-e}.

\section{Conclusions and discussion}
In the present work we propose a new initial state model for hydrodynamic simulation of relativistic heavy ion collisions based on Bjorken-like solutions applied streak by streak in the transverse plane and producing an IS qualitatively similar to the results of parton cascade models like \cite{Fig.1}. Our IS can be given in both [t,x,y,z] and $[\tau, x, y, \eta]$ coordinates, and thus can be tested by all 3+1D hydrodynamical codes which exist in the field. Most importantly, it is able to incorporate
initial shear, in contrary to several other initial state 
parametrizations. The lack of initial shear reduces 
the vorticity and the possibility for polarization in those
models, which contradicts recent observations \cite{Lisa,Nature}.

The  velocity  distributions produced in our initial state model are shown in Figs. \ref{finvx}, \ref{vxdy} and \ref{Tauf-v-e} in the reaction plane, $y=0$.  As we see among the different streaks of the matter there is considerable shear, particularly for peripheral
streaks, {\it e.g.} $y$ = 4fm, see Fig. \ref{vxdy}. Fig. \ref{Tauf-v-e}
indicates that the velocity profile shows dominant longitudinal
expansion, which gradually may decrease the central shear. Thus, the 
development of a Kelvin-Helmholtz Instability   in this configuration is less probable than  in earlier calculations with a different initial state \cite{hydro2}.

The shear will lead to strong vorticity. This vorticity vector pointing in the $-y$ direction, dominates the vorticities developing due to
the expansion later in the flow. Furthermore, due to symmetry
reasons the vorticities in the other directions cancel each other
to a large extent \cite{Xie2016,Xie2017}, except for eventual unbalanced
vorticities due to random fluctuations
\cite{Floe2013,Stefan}.

The energy weighted classical vorticity, $\omega_y(x,\eta)$, see \cite{CMW13} for detailed definition, is
shown in Fig. \ref{vort-ew_y}. This component is overall negative
arising from the initial rotation,
{\it i.e.}, it is pointing in the $-y$ direction. 
The central part of the momentum domain
at this initial moment shows smaller vorticity, due to the
Bjorken-like expansion of the model.

\begin{figure}[htb]     
\begin{center}
\resizebox{1.01\columnwidth}{!}
{\includegraphics{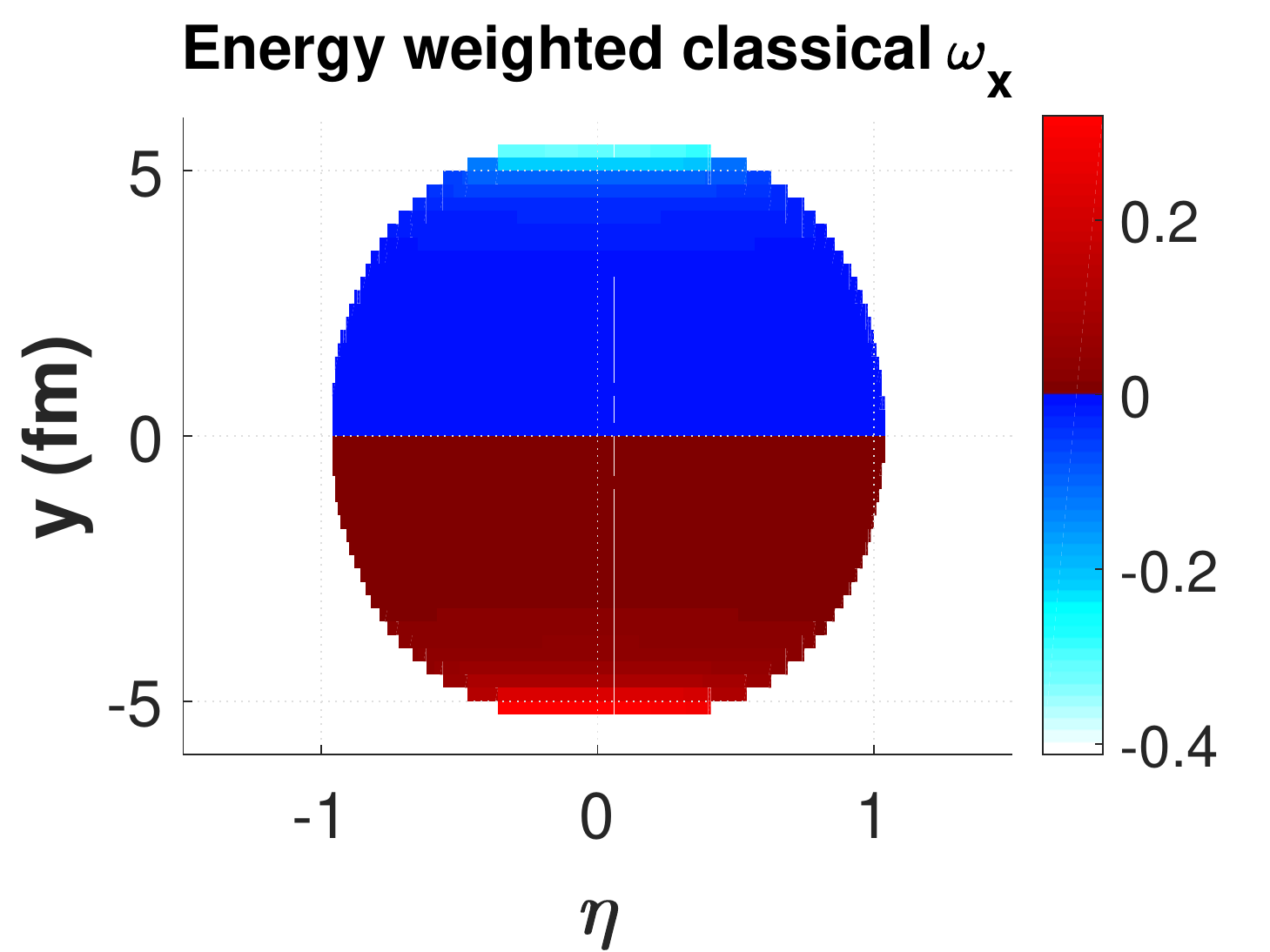}}
\caption{ (color online)
The $x$-component of the classical vorticity
on the $[y,\eta]$ plane at $x=3.0$ fm. This component of vorticity,
has similar values as the $y$ component, but it is 
antisymmetric, 
$\omega_x(y_{max}, \eta=0) = - \omega_x(y_{min}, \eta=0)$.
So the two identical but opposite signed vorticities
yield a vanishing overall sum.
}
\label{vort_x}
\end{center}
\end{figure}

The classical vorticity, $\omega_x(y,\eta)$ is
shown in Fig. \ref{vort_x}. This component is antisymmetric 
across the $y=0$ surface.
As a consequence the contribution of this component 
vanishes in the complete averaging.
The central part of the domain
at this initial moment shows smaller vorticity.

The vorticity is observed via the observed polarization, $\Pi_y$, of emitted 
$\Lambda$ and $\bar{\Lambda}$ particles 
\cite{Bec13,Erratum,BCDG13}. The symmetries of the vorticity field and
of the momentum dependence of the polarizations are tightly
related \cite{BCDG13}. The present experiments show only the overall
$\Lambda$ and $\bar{\Lambda}$ polarizations summed up for all emission
momenta. Thus, in the c.m. frame the polarization components,
$\Pi_x$  and $\Pi_z$ must vanish, 
due to the symmetries of vorticity components,
$\omega_x$ and $\omega_z$,
except a smaller contribution from random fluctuations.

At the same time the $x$ and $y$ vorticity components carry valuable
information, but these can only be extracted if the participant 
c.m. is identified Event by Event (EbE)
\cite{Eyyubova,CseStoe2014}.
This identification based on the spectators detected via the
zero degree calorimeters is not performed yet experimentally
\cite{Schukraft},
due to assumed, unrelated fluctuations of other origins.
Now this identification method could be tested by evaluating the
sum of polarizations, $\Pi_x$  and $\Pi_z$, with and without
EbE identification of participant c.m. 
With c.m. identification the $x$ and $y$ polarization components
should vanish or become minimal. The
$\Pi_x(\vec{p})$  and $\Pi_z(\vec{p})$ distributions 
should also show the 
symmetries arising from the symmetries of the vorticities.
This will provide valuable information on the details of the
initial state models which cannot be easily detected in other 
ways.

The current model is a simple realization for peripheral
heavy-ion collisions, with initial shear and vorticity,
in Milne coordinates. Unlike the large majority of the 
Bjorken type of models that do not discuss the longitudinal
degrees of freedom, we divide the transverse plane to streaks
that are longitudinally finite. At every transverse
point $i \equiv [x_i,y_i]$ we have a longitudinal streak 
with well defined end points, 
$z^{max}_{i,P}$ and $z^{min}_{i,P}$ or the corresponding
points in Milne coordinates
$\eta^{max}_{i,P}$ and $\eta^{min}_{i,P}$ on the Projectile
side. We describe the Target side similarly. We obtain these
points from the streak by streak energy and momentum conservation,
and from simple assumptions regarding the streak ends and
streak center points.
  
There exist a few models in Milne coordinates, which  discuss 
the longitudinal degrees of freedom in the collisions, and satisfy 
energy and momentum conservations. For example 
ref. \cite{Mishustin2011} introduces streak ends,
$z_p(\tau)$ and $z_t(\tau)$, but these are uniform,
{\it i.e.}, identical for all transverse points. This model could 
be generalized in the same way to varying peripheral streaks,
so that energy and momentum conservation is applied streak
by streak, and as a consequence shear and vorticity will
be included in the model. In this case every transverse
streak would have a different constant proper time
hyperbola, with different origins in the space-time
$[t_{i0}, z_{i0}]$.

A very interesting approach has been presented at Quark Matter 2017 Conference \cite{Shen:2017ruz}. The whole model is still in preparation, but according to the Figs. in the Proceedings \cite{Shen:2017ruz} this model also has asymmetric hyperbolae, which appear as "thermalized strings". As far as we understood the asymmetry in this model is related to IS fluctuations in the position/time of the initial parton collisions, and it is not clear whether it is systematically increasing with $x$, as in our case.  Also we would like to note that the model of \cite{Shen:2017ruz} has zero pressure free streaming before thermalization, which may lead to unrealistically increasing
  transparency and may eliminate the development of local vorticity
  in peripheral collisions.  This is in stark contrast to the
  field dominant initial state dynamics described in
\cite{M2001,M2002,Gy1986,MiL2012,Stoecker:2015zea}.

In case of Color Glass Condensate in the initial state, the 
colour field slows down the leading charges of the 
expanding system, as discussed in 
\cite{FKL16,OF14,CFKL15,Fries:2017ina}.
One can follow the trajectory of the longitudinal edges up 
to some $\tau = const.$ hypersurface, and obtain the corresponding space-time rapidities 
$\eta_{min}$ and $\eta_{max}$, which limit the longitudinal 
extent of the flux-tube with the gluon field or plasma. The field
may even contribute to a large compression of the baryon charge
at the forward and backward edges
\cite{LK17,Gy1986}.
This model could also 
be generalized to varying peripheral streaks,
so that energy and momentum conservation is applied streak
by streak and the streak ends 
would be different for each peripheral streak.

And finally we would like to mention that by varying the parameters of our model, namely $\Delta \eta_c$, $\tau_0$, $t_{IS}$ or $\tau_{IS}$, the geometry of the produced IS can be adjusted to the different parton cascade approaches as well as to the different field theoretical models. For example, using parameter set $\Delta \eta_c=1.7$, $\tau_0=2.0$ fm/c, $t_{IS}=3.2$ fm/c  we managed to reproduce rather closely the form and volume of the IS from Ref. \cite{M2002} (of course, the flow velocity and energy density distributions are still fixed by the Bjorken nature of the model and stay rather different from those in \cite{M2002}). This feature of the proposed approach may provide a basis for further studies of different 
physical processes.

\begin{acknowledgments}

Enlightening discussions with Hannah Petersen and Etele Molnar
are gratefully acknowledged.
One of the authors, Y.L. Xie 
is supported by China Scholarship Council (China) 
and by COST Action CA 15213. 
This work was partially supported by 
by the Alexander von Humboldt Foundation,
by the Research Council of Norway, Grant \# 255253/F50, 
by the Spanish Ministerio de Economia y Competitividad (MINECO) under the project MDM-2014-0369 of ICCUB (Unidad de Excelencia 'Mar\'\i a de Maeztu'), Ge\-ne\-ra\-li\-tat de Catalunya, contract 2014SGR-401,
and, with additional European FEDER funds, under the contract FIS2014-54762-P and the Spanish Excellence Network on Hadronic Physics FIS2014-57026-REDT.

\end{acknowledgments}

\section*{APPENDIX: Propagation to the central streak frame}

\begin{figure}[htb]     
\begin{center}
\resizebox{1.01\columnwidth}{!}
{\includegraphics{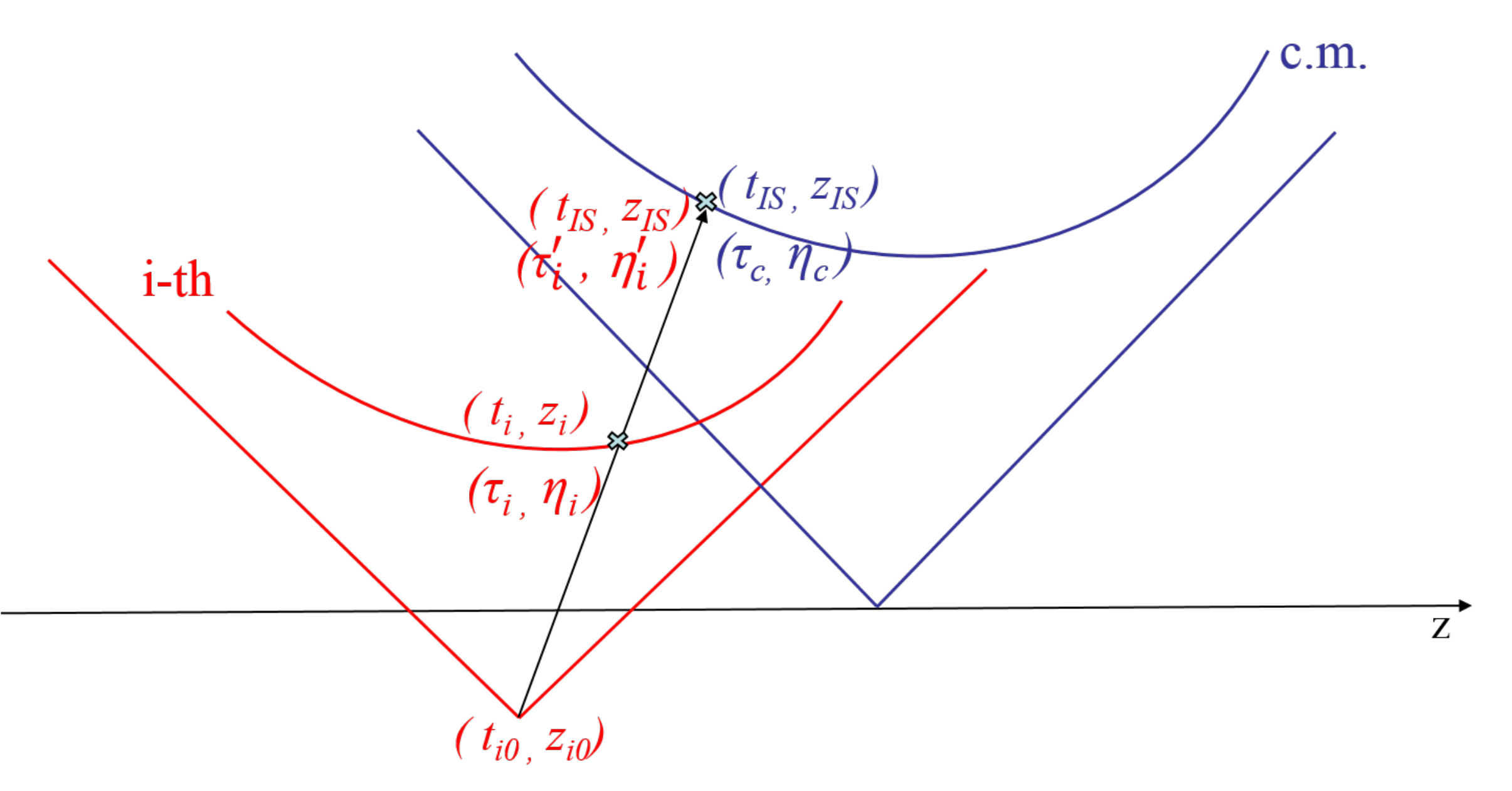}}
\caption{
The propagation from the $i$-th peripheral hyperbola to the
central streak hyperbola, in order that all streaks will be situated 
on a joint (blue) hyperbola at $\tau_c =\tau_{IS} = const.$
The primary situation of the initial state is on the (red)
$i$-th peripheral hyperbola at $(t_i, z_i)$ or $(\tau_i, \eta_i)$.
With the same rapidity, $\eta_i$, this fluid element is propagated
to the (blue) c.m. hyperbola to point 
$(t_{IS}, z_{IS})$ or $(\tau_{i} ', \eta_i ')$.
The same point in the c.m. frame is on the (blue) central streak hyperbola
at the same space-time coordinates $(t_{IS}, z_{IS})$, but in the
(blue) frame's coordinates, it is $(\tau_{c}, \eta_c)$.
}
\label{propag}
\end{center}
\end{figure}        

As we have discussed in Section \ref{Milne}, after Bjorken expansion, 
the central streak and $i$-th peripheral streak, with different initial 
points,  
will stop at the $\tau_c = \tau_i = \tau_{IS}$ hyperbolae, which are shown 
in Fig. \ref{propag}. Now we map the solution of $\tau_i = \tau_{IS}$ to 
the $\tau_c = \tau_{IS}$ hyperbola, 
which we call `propagation', by keeping the 
rapidity $\eta_i$ unchanged. 
{\it i.e.}, from Eq. (\ref{eq44}) the point $(t_i, z_i)$ in Fig.\ref{propag}, with its propagated point $(t_{IS}, z_{IS})$ will have the same 
rapidity in the $i$-th peripheral streak's frame:
\be
\eta_i = \eta_i ' =
      \Artanh\left(\frac{\tau_{IS} \sinh \eta_c - z_{i0}}{\tau_{IS} \cosh \eta_c - t_{i0}}	
\right) \,.
\label{eq44-2}
\ee

This equation combined with eqs. (\ref{deftz},\ref{eq43}) describe the propagation, see Fig. \ref{propag}.

Now from assumption (a), we have limits on $z$ coordinate of the $i$-th
streak: $z_{max}^{i,P} = z_{max}^c$ and $z_{min}^{i,T} = z_{min}^c$, which results on limits on rapidity, {\it i.e.},  
$\eta_{min}^i\le \eta_i \le \eta_{max}^i$.
Imposing these conditions on the $\eta_i '$, given by eq. (\ref{eq44-2}), we can find the corresponding limits for the $\eta_c$ for the $i$-th streak. 

Thus, 
\be
\tanh \eta_i' = 
\left(\frac{\tau_{IS} \sinh \eta_c -z_{i0}}
{\tau_{IS} \cosh \eta_c - t_{i0}}\right)
\le \tanh \eta_{max}^{i} ,
\label{thetam}
\ee
where $\eta_{max}^{i}=\, <\eta_i> + \frac{\Delta\eta_i}{2}$, calculated as explained in section \ref{sec3}. 
Multiplying the second inequality by
$\left[\cosh \eta_{max}^{i} \cdot (\tau \cosh \eta_c-t_{i0})\right]$,
we get 
\be
(\tau_{IS} \sinh \eta_c -z_{i0}) \cosh \eta_{max}^{i}  \le
(\tau_{IS} \cosh \eta_c -t_{i0}) \sinh \eta_{max}^{i} \,.
\ee

Performing the multiplications, and using the expression 
of $\sinh(A-B)$ we obtain
\be
\sinh(\eta_c - \eta_{max}^{i}) \le \frac
{z_{i0} \cosh \eta_{max}^{i} - t_{i0} \sinh \eta_{max}^{i}}{\tau_{IS}}\,.
\ee
Now taking $\Arsinh$ of this equation leads to eq. (\ref{eq48}):
$$
\eta_c  \le \eta_{max}^{i} + \Arsinh \left(\frac
{z_{i0} \cosh \eta_{max}^{i} - t_{i0} \sinh \eta_{max}^{i}}{\tau_{IS}}
\right) \,.
$$
Similarly one can get eq. (\ref{eq49}).


\end{document}